\documentclass[openany]{aa}
\usepackage[varg]{txfonts}
\usepackage[normalem]{ulem}
\usepackage{graphicx}
\usepackage{color}
\usepackage{float}
\graphicspath{ {./plots/} }
\usepackage{natbib}
\bibpunct{(}{)}{;}{a}{}{,}
\usepackage{multicol} 
\usepackage{array,multirow}
\usepackage{footnote}
\usepackage{tablefootnote}

\usepackage{url}
\usepackage[titletoc]{appendix}
\usepackage{tabularx}
\usepackage{soul}
\usepackage{babel,blindtext}
\usepackage{xcolor}
\usepackage{hyperref}
\hypersetup{colorlinks, allcolors = {blue!50!black}}
\usepackage{cleveref}
\usepackage{subfigure}
\usepackage{orcidlink}
\usepackage{placeins}

\definecolor{tabblue}{HTML}{1F77B4}
\definecolor{taborange}{HTML}{FF7F0E}
\definecolor{tabgreen}{HTML}{2CA02C}
\definecolor{tabred}{HTML}{D62728}
\definecolor{tabpurple}{HTML}{9467BD}
\definecolor{tabbrown}{HTML}{8C564B}
\definecolor{tabpink}{HTML}{E377C2}
\definecolor{tabgray}{HTML}{7F7F7F}
\definecolor{tabolive}{HTML}{BCBD22}
\definecolor{tabcyan}{HTML}{17BECF}

\definecolor{redorange}{rgb}{2.50, 0.7, 0.22}

\begin{document} 

\title{Revealing a ribbon-like jet in OJ 287 with RadioAstron}

\author{
E.~Traianou\inst{\ref{iaa},\ref{iwr}} \orcidlink{0000-0002-1209-6500} \and 
J. L.~G\'omez\inst{\ref{iaa}} \orcidlink{0000-0003-4190-7613} \and
I.~Cho\inst{\ref{kasi}, \ref{yonsei}, \ref{iaa}} \orcidlink{0000-0001-6083-7521} \and 
A.~Chael \inst{\ref{pgi}}\orcidlink{0000-0003-2966-6220} \and 
A.~Fuentes\inst{\ref{iaa}} \orcidlink{0000-0002-8773-4933} \and I.~Myserlis\inst{\ref{iram}, \ref{mpifr}} \orcidlink{0000-0003-3025-9497} \and 
M.~Wielgus\inst{\ref{iaa}} \orcidlink{0000-0002-8635-4242} \and 
G.-Y.~Zhao\inst{\ref{mpifr}, \ref{iaa}} \orcidlink{0000-0002-4417-1659} \and 
R.~Lico\inst{\ref{inaf}} \orcidlink{0000-0001-7361-2460} \and 
K.~Moriyama\inst{\ref{iaa}} \orcidlink{0000-0003-1364-3761} \and 
L.~Dey\inst{\ref{wvu},\ref{gwac}} \orcidlink{0000-0002-2554-0674} \and 
G.~Bruni \inst{\ref{iaps}} \orcidlink{0000-0002-5182-6289} \and  
R.~Dahale\inst{\ref{iaa}} \orcidlink{0000-0001-6982-9034} \and 
T.~Toscano\inst{\ref{iaa}} \orcidlink{0000-0003-3658-7862} \and 
L.~I.~Gurvits\inst{\ref{jive}, \ref{tud}, \ref{shao}} \orcidlink{0000-0002-0694-2459} \and 
M.~M.~Lisakov\inst{\ref{PUCV}} \orcidlink{0000-0001-6088-3819} \and 
Y.~Y.~Kovalev\inst{\ref{mpifr}} \orcidlink{0000-0001-9303-3263} \and 
A.~P.~Lobanov\inst{\ref{mpifr}} \orcidlink{0000-0002-1209-6500} \and
A.~B.~Pushkarev\inst{\ref{crimea}, \ref{moscow}} \orcidlink{0000-0002-9702-2307} \and
K.~V.~Sokolovsky\inst{\ref{uiuc}} \orcidlink{0000-0001-5991-6863}
}

\institute{Instituto de Astrof\'\i sica de Andaluc\' \i a (IAA-CSIC), Glorieta de la Astronom\' \i a s/n, 18008 Granada, Spain \label{iaa} \\
\email{traianouthalia@gmail.com}
\and Interdisziplin\"ares Zentrum f\"ur Wissenschaftliches Rechnen (IWR), Universit\"at Heidelberg, Im Neuenheimer Feld 205, 69120 Heidelberg, Germany \label{iwr}
\and Korea Astronomy and Space Science Institute, Daedeok-daero 776, Yuseong-gu, Daejeon 34055, Republic of Korea \label{kasi}
\and Department of Astronomy, Yonsei University, Yonsei-ro 50, Seodaemun-gu, 03722 Seoul, Republic of Korea \label{yonsei}
\and Max-Planck-Institut f\"ur Radioastronomie, Auf dem H\"ugel 69, D-53121, Bonn, Germany \label{mpifr}
\and Princeton Gravity Initiative, Princeton University, Princeton NJ, 08540 \label{pgi}
\and INAF $-$ Istituto di Radioastronomia, via Gobetti 101, 40129 Bologna, Italy \label{inaf} 
\and INAF $-$ Istituto di Astrofisica e Planetologia Spaziali, via del Fosso del Cavaliere 100, Roma, 00133, Italy \label{iaps}
\and Institut de Radioastronomie Millim\'etrique, Avenida Divina Pastora, 7, Local 20, E18012 Granada, Spain \label{iram}
\and Department of Physics and Astronomy, West Virginia University, P.O. Box 6315, Morgantown, WV 26506, USA \label{wvu}
\and Center for Gravitational Waves and Cosmology, West Virginia University, Chestnut Ridge Research Building, Morgantown, WV 26505, USA \label{gwac}
\and Joint Institute for VLBI ERIC (JIVE), Oude Hoogeveensedijk 4, 7991 PD Dwingeloo, The Netherlands \label{jive}
\and Faculty of Aerospace Engineering, Delft University of Technology, Kluyverweg 1, 2629 HS Delft, The Netherlands \label{tud}
\and Shanghai Astronomical Observatory, Chinese Academy of Sciences, 80 Nandan Rd., Shanghai 200030, China \label{shao}
\and Instituto de F\'{i}sica, Pontificia Universidad Cat\'{o}lica de Valpara\'{i}so, Casilla 4059, Valpara\'{i}so, Chile \label{PUCV} 
\and Crimean Astrophysical Observatory, 298409 Nauchny, Crimea \label{crimea}
\and Institute for Nuclear Research of the Russian Academy of Sciences, 60th October Anniversary Prospect 7a, Moscow 117312, Russia \label{moscow}
\and Department of Astronomy, University of Illinois Urbana-Champaign, 1002 W. Green Street, Urbana, IL 61801, USA \label{uiuc} \\
}

   \date{Version: \today; Received \dots; accepted \dots}

\abstract{
We present space-based very long baseline interferometry observations of the BL\,Lac type object \object{OJ\,287} taken with RadioAstron at 22\,GHz on April 25, 2016, in conjunction with a ground array comprising 27 radio telescopes. 
We detect ground-space fringes at projected baselines extending up to 4.6 Earth diameters, which allowed us to image the jet in \object{OJ\,287} with an angular resolution of $\sim47\,\mu$as. Applying an advanced regularized maximum likelihood imaging method, we resolved the innermost jet structure with a complex morphology at a resolution of $\sim$15$\,\mu$as ($\sim$0.1\,pc projected distance). For the first time, due to a favorable geometrical position of the jet in tandem with high data quality, we detect multiple sharp bends that form a ``ribbon-like'' jet structure that extends down to 1 mas. Two-dimensional Gaussian model-fitting reveals regions of the jet with brightness temperatures of more than $10^{13}$\,K, indicative of strong Doppler boosting. Polarimetric imaging reveals that the electric vector position angles are predominantly perpendicular to the innermost jet direction, implying a dominant poloidal magnetic field component near the central engine. Complementary multi-epoch Very Long Baseline Array observations at 43\,GHz provide a multifrequency view of the jet evolution. Ridgeline analysis of the 43\,GHz data shows significant variations in the jet position angle from 2014 to 2017, behavior consistent with a rotating helical jet structure. Finally, we confirm the emergence of a new jet component (B15 or K), which may be associated with the source's first TeV flare, and offer new observational constraints relevant to models involving a supermassive black hole binary.} 


\maketitle
\makeatletter
\let\linenumbers\relax
\let\endlinenumbers\relax
\let\do@mlinenumbers\relax
\let\@LN@do@ln\relax
\let\makeLineNumber\relax
\def\linenumberfont{} 
\renewcommand\thelinenumber{} 
\setlength\linenumbersep{0pt}
\setlength\linenumberwidth{0pt}
\makeatother


\section{Introduction}
\label{sec:intro}

Among the various classes of active galactic nuclei (AGNs), BL Lacertae (BL\,Lac) objects stand out for their rapid, large-amplitude variability and significant polarization across multiple wavebands, which is attributed to relativistic jets aligned closely with our line of sight \citep[e.g.,][]{2017NatAs...1E.194P}. A prominent member of this subclass is \object{OJ\,287}, situated at a redshift of $z=0.306$ \citep{Stickel+1989}. Optical observations of \object{OJ\,287} extend back to the 1880s \citep{Sillanpaa1988}, resulting in an exceptionally long light curve spanning nearly 150 years. This extensive dataset reveals quasiperiodic brightness fluctuations, including a prominent $\sim$60-year cycle \citep{Valtonen2006} and recurrent, doubly peaked high-luminosity flares approximately every 12 years \citep{Valtonen2006,Dey2018}. These periodic variations are explained well by a supermassive black hole binary (SMBHB) model in which a secondary supermassive black hole follows a precessing, eccentric orbit around a more massive primary. Flares are generated each time a smaller component crosses the primary's accretion disk \citep{LehtoValtonen1996,Sundelius1997,Dey2019}. Within this framework, the well-known ``impact outbursts'' of \object{OJ\,287} \citep[e.g.,][]{Valtonen2008Nat} are the direct manifestations of these disk crossings.

In recent years, \object{OJ\,287} has gained significant attention due to its potential connection to the long-anticipated direct gravitational wave (GW) detections by the LIGO-VIRGO collaboration in 2015 \citep{Abbott2016GW-disc}. The following reports of a GW background from pulsar timing arrays such as NANOGrav \citep{Agazie2023NANOGrav} have reinforced the idea that SMBHBs are primary contributors to the GW background in the range of frequencies probed with pulsar timing. Also, \object{OJ\,287} proved to be a ``testing ground'' for multi-messenger studies based on electromagnetic and GW observations even before the first GW detections \citep[see, e.g.,][]{Valtonen2008Nat}. (Sub)milliarcsecond-scale very long baseline interferometry (VLBI) images of \object{OJ\,287} provide crucial information for determining the stage of SMBHB evolution along the track defined by \cite{Begelman1980Nat}.

The parsec-scale jet of \object{OJ\,287} shows interesting behavior too; for example, it undergoes dramatic position angle (PA) variations of up to $\sim$130$^\circ$ \citep{Agudo2012,Cohen2017}, which frequently coincide with major flaring episodes. These reorientations likely arise from a combination of geometric effects (such as jet precession) and dynamical processes, including the propagation of powerful shocks. Several mechanisms could explain these PA variations, including plasma instabilities and frame-dragging from a jet--disk misalignment, similar to patterns seen in other AGN jets \citep{Abraham2000,CaproniAbraham2004,TateyamaKingham2004,Britzen2018,Qian2018}. 
Past VLBI observations have connected these PA rotations with structural changes in the jet, including component ejections, core flux variations, and transient features in the innermost regions \citep[e.g.,][]{Agudo2012,2017A&A...597A..80H}. However, attempts to model the high-frequency radio behavior using purely geometric or instability-driven approaches have yielded inconsistent results, as these models often fail to account for the complex, multi-zone nature of relativistic jets. For instance, \citet{Agudo2012} demonstrated that single-zone models cannot reproduce the observed rapid polarization variability, suggesting the necessity for multi-zone models to explain the simultaneous occurrence of polarization angle swings and flux density flares. Similarly, \citet{2017A&A...597A..80H} find that purely geometric models, such as those involving helical magnetic fields, are insufficient to explain the observed polarization variability, indicating that additional physical processes, like shock propagation or magnetic reconnection, must be considered to accurately model the polarization behavior in AGN jets. Additionally, the failure to detect the predicted October 2022 outburst \citep{Komossa2023_Impact} posed a significant challenge to certain SMBHB models. In response, refinements to these frameworks have incorporated additional factors, such as the geometry of the accretion disk, to improve predictive accuracy \citep{Valtonen2023}.

Space VLBI offers unprecedented angular resolution on the order of tens of microarcseconds, enabling detailed studies of AGN jets in total and linearly polarized intensity \citep[e.g.,][]{Fuentes2023}. The 10-meter space radio telescope (SRT) RadioAstron \citep{2013ARep...57..153K} operated from 2011 to 2019. It served as the primary instrument aboard the {Spektr-R} spacecraft, which functioned as its orbital platform. RadioAstron conducted observations at frequencies ranging from 1.6 to 22\,GHz, and its orbit extended to $\sim$350,000\,km, providing exceptional angular resolution at its shortest wavelengths. Results from the RadioAstron polarization key science program have significantly advanced our understanding of jet physics. The program encompassed diverse source classes, including
BL\,Lac objects \citep{Gomez_2016}, 0716$+$714 \citep{2020ApJ...893...68K}, 0642$+$449 \citep{2015A&A...583A.100L}, 3C\,345 \citep{2021A&A...648A..82P}, 3C\,273 \citep{2017A&A...604A.111B}, and more.

Dedicated observations of \object{OJ\,287} with RadioAstron at 22\,GHz from April 4-5, 2014, achieved an unprecedented angular resolution of 12\,$\mu$as, revealing a progressively bending jet extending down to $\sim$0.3 mas, providing key evidence of a predominantly poloidal magnetic field in the core, and suggesting that the parsec-scale jet maintains equipartition between particles and magnetic field energies \citep{2022ApJ...924..122G}. Close in time, 1.68\,GHz RadioAstron observations probed larger jet scales, resolving structures of up to $\sim$10\,pc in size \citep{2024A&A...683A.248C} and confirming a similar energetic equilibrium. Meanwhile, Global Millimeter VLBI Array (GMVA) observations at 3.5\,mm (86\,GHz) in April 2017, combined with new imaging techniques such as regularized maximum likelihood (RML), revealed a twisted, complex inner jet structure and polarization signatures indicative of recollimation shocks and a helical magnetic field configuration \citep{2022ApJ...932...72Z}. In the same year, the Event Horizon Telescope observed the source for the first time at 1\,mm \citep{Gomez2025}. Nevertheless, for all these observations, at $\sim$20\,$\mu$as resolution the source showed a jet extending only down to 0.2-0.3\,mas.

In this paper we extend previous studies by presenting RadioAstron observations of \object{OJ\,287} at 22\,GHz from April 26, 2016. For the first time, we image a continuous and sharply bent jet morphology down to 1\,mas, enabled by the combination of high sensitivity, high angular resolution, and a favorable jet orientation. Although previous high-resolution VLBI studies \citep[e.g.,][]{Agudo2012,2017A&A...597A..80H} have resolved features within this scale, the multiband, ribbon-like structure revealed here had remained undetected, probably due to a limited dynamic range or a less optimal viewing geometry. 

By juxtaposing high-resolution space VLBI images with quasi-simultaneous multiwavelength data, we investigate how the jet morphology and polarization structure have evolved.

\section{Observations and data analysis} 
\label{sec:obs_im}

\begin{figure*}
\centering
\includegraphics[scale=0.11]{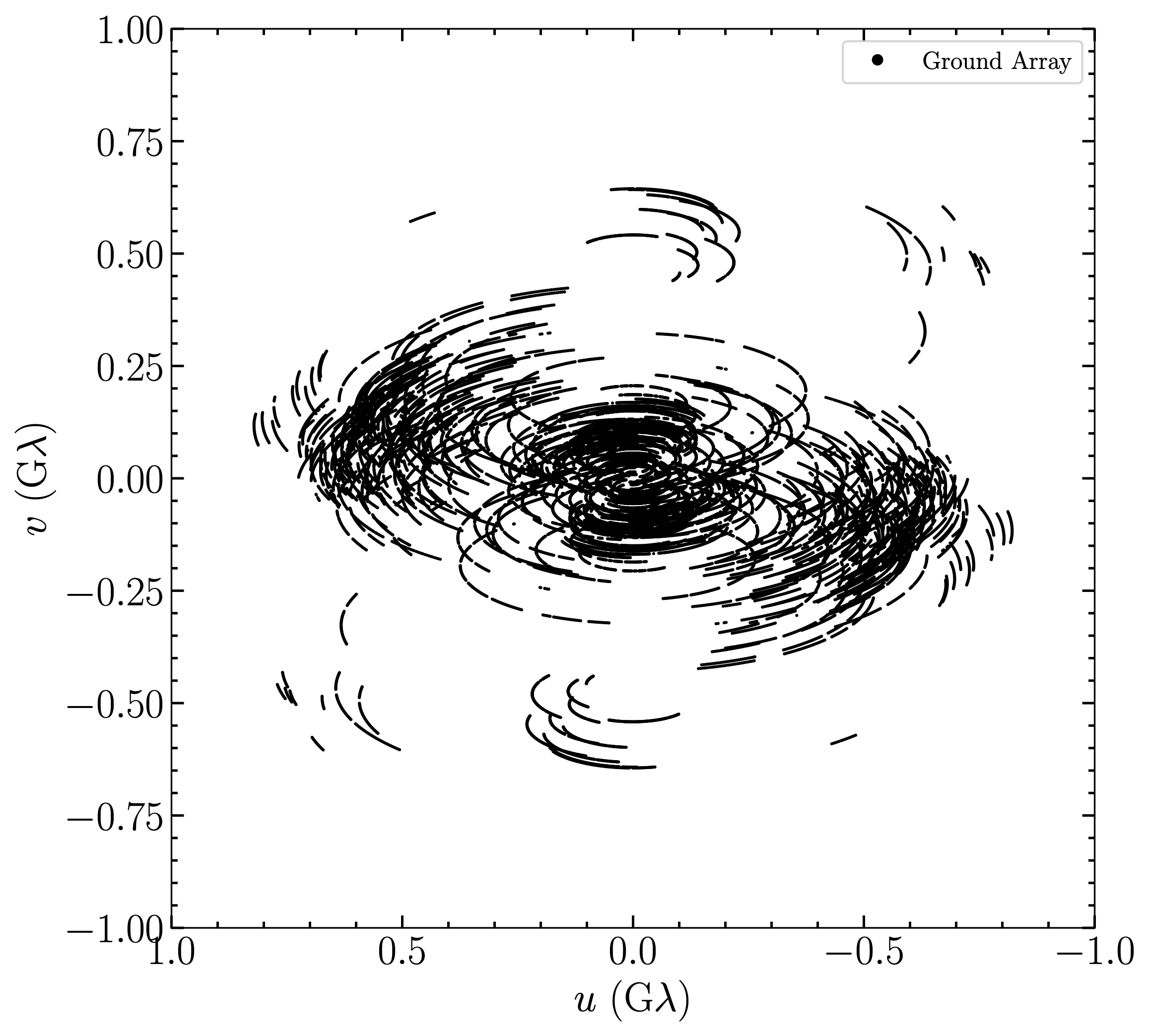} \hspace{0.5cm}
\includegraphics[scale=0.11]{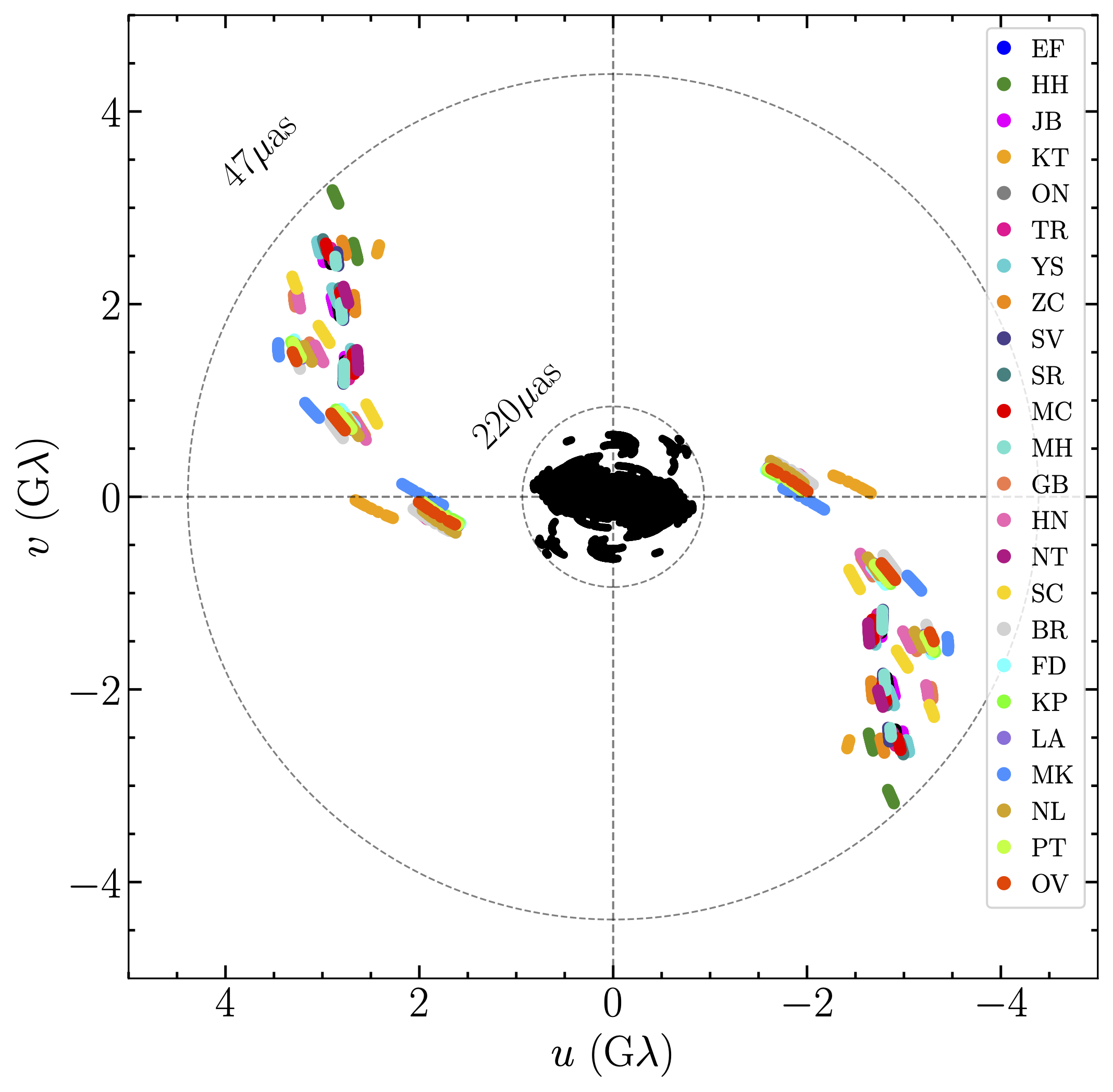}
  \caption{($u,v$) coverage of our observation at 22\,GHz. \textit{Left:} Ground-only array. \textit{Right:} Full observations with ground-based arrays plus the SRT. Different colors show the ground stations providing the baseline with the SRT (see Table \ref{tab:stations} for station codes). The maximum baseline lengths of the space baseline and the ground baseline are shown with dashed circles, which correspond to angular resolutions of 47\,$\mu$as and 220\,$\mu$as, respectively. Each point has been averaged with a 2-minute interval in both panels.
  }
     \label{fig:uv}
\end{figure*}

The RadioAstron mission observed \object{OJ\,287} at 22.24 GHz ($\lambda = 1.35$ cm) on April 25, 2016, from 16:00 to April 26 at 06:50~UT, 
spanning a total of approximately 15 hours,  including about 12 hours on source (experiment code GG079C). The observations included calibrators 0716$+$714, 3C~345, 0059$+$581, 4C~$+$38.41, and 4C~$+$45.51 
for ground-based radio telescopes.
A total of 27 ground-based telescopes and the SRT participated in the experiment, providing extensive u-v coverage (see Table~\ref{tab:stations} and Fig.~\ref{fig:uv}). 
The data were recorded in both left (LCP) and right (RCP) circular polarizations, with a total bandwidth of 64\,MHz per polarization (512~Mbps with 2-bit sampling), 
split into four         intermediate frequency (IF) bands for the ground antennas. For the SRT, the bandwidth was 32\,MHz per polarization, 
split into two IFs and sampled at 1-bit. Due to operational issues, data from the Badary (BD), Ulsan (KU), and Yonsei (KY) stations were excluded from 
the analysis.

The SRT data were downlinked in real time to the RadioAstron tracking stations in Pushchino and Green Bank, with three-hour gaps to cool the onboard data downlink system of the Spektr-R satellite. The data were processed using the RadioAstron-dedicated version of the \texttt{DiFX} software correlator \citep{2016Galax...4...55B}, 
developed at the Max-Planck-Institut f\"ur Radioastronomie. The SRT provided baselines longer than the Earth's diameter ($D_\oplus$), resulting in a $u$-$v$ extension up to 4.3\,G$\lambda$ (4.6\,$D_\oplus$, Fig.~\ref{fig:uv}), achieving an angular resolution of $\sim$47\,$\mu$as. The spacecraft's highly elongated orbit in the east--west direction yields higher resolution along the east-southeast to west-northwest direction.

The calibration of the RadioAstron data was performed using NRAO's \texttt{AIPS} software package \citep{2003ASSL..285..109G}, following similar procedures to those described in \cite{2022ApJ...924..122G} and \cite{2024A&A...683A.248C}. The process involved solving for residual single- and multi-band delays, phases, and phase rates through gradual fringe fitting of the data. In the first iteration, RadioAstron was excluded, and a global fringe search was performed on the ground array with a solution interval of 60\,s, with a signal-to-noise ratio (S/N) threshold of 5. In subsequent fringe fitting that included the SRT, a lower S/N threshold of 4 was used for space–ground baselines to account for their typically lower fringe amplitudes. Effelsberg and the Green Bank Telescope were used as reference antennas throughout the process.

Manual phase calibration was done iteratively, starting with the ground array and then progressively incorporating the RadioAstron SRT. Fringe fitting was performed between the ground telescopes and the SRT in each segment separately. 
The ground array was then coherently combined through baseline stacking to increase the S/N of possible fringe detections to the SRT \citep[e.g.,][]{2017A&A...604A.111B}. 

To account for the spacecraft's residual acceleration near perigee, we adopted different solution intervals ranging from 10 to 180 seconds during fringe fitting. 
While the acceleration term was checked during correlation, it was not applied, as it proved unnecessary in practice. On shorter baselines, the higher S/N permitted shorter solution intervals, making acceleration corrections straightforward in post-correlation analysis. On longer baselines, the effect of acceleration was negligible and did not require correction. Additionally, since the longest projected baselines were less sensitive to acceleration, we optimized the total data bandwidth by combining IFs. Polarization alignment was performed using \texttt{RLDLY} task on a bright scan of the source. Amplitude calibration was performed by loading and applying the system temperature and gain curve information for each antenna, and deriving amplitude corrections using \texttt{APCAL}, while accounting for issues at specific antennas. The amplitude solutions were smoothed using \texttt{SNSMO}. Due to the low S/N of many scans, instead of performing a bandpass calibration, the outer six spectral channels on each side of the IFs were flagged to eliminate signal attenuation and leakage. 

To complement our RadioAstron data, we also incorporated archival 43\,GHz Very Long Baseline Array (VLBA) observations to study the jet morphology over time. 
The 43\,GHz data used in this work (presented in Fig.~\ref{fig:bu}) were obtained with the VLBA, spanning the period from 2014 to 2017 with annual cadence, within the VLBA-BU-BLAZAR program\footnote{\url{https://www.bu.edu/blazars/BEAM-ME.html}}. 
The program conducts regular monthly observations of a sample of $\gamma$-ray bright AGNs. A detailed description of the observations and data reduction can be found in \cite{2017ApJ...846...98J} and \cite{2022ApJS..260...12W}. For this study, we reimaged a total of four images.

\subsection{Imaging and model-fitting: Total intensity}

\begin{table}
\centering
\small
\caption{Observing stations.}
{\small
\setlength{\tabcolsep}{4pt} 
\begin{tabular}{llcc}
\hline 
Array                  & Station        & Code & Diameter \\
 & & & (m) \\
\hline 
\multirow{10}{*}{VLBA} & Brewster       & BR   & 25           \\
                       & Fort Davis     & FD   & 25           \\
                       & Hancock        & HN   & 25           \\
                       & Kitt Peak      & KP   & 25           \\
                       & Los Alamos     & LA   & 25           \\
                       & Mauna Kea      & MK   & 25           \\
                       & North Liberty  & NL   & 25           \\
                       & Owens Valley   & OV   & 25           \\
                       & Pie Town       & PT   & 25           \\
                       & Saint Croix    & SC   & 25           \\
                       \\
NRAO                   & Green Bank     & GB   & 100          \\
\\
\multirow{13}{*}{EVN}  & Effelsberg     & EF   & 100          \\
                       & Hartebeesthoek & HH   & 26           \\
                       & Jodrell Bank   & JB   & 25           \\
                       & Medicina       & MC   & 32           \\
                       & Robledo        & RO   & 70           \\
                       & Toru\'{n}      & TR   & 32           \\
                       & Yebes          & YS   & 40           \\
                       & Noto           & NT   & 32           \\
                       & Onsala         & O6   & 25           \\
                       & Mets\"{a}hovi  & MH   & 14           \\
                       & Svetloe        & SV   & 32           \\
                       & Zelenchukskaya & ZC   & 32           \\
                       & Badary         & BD   & 32           \\
\\                       
\multirow{3}{*}{KVN}   & Yonsei         & KY   & 21           \\
                       & Ulsan          & KU   & 21           \\
                       & Tamna          & KT   & 21           \\
\\
SRT                    & Spektr-R       & RA   & 10 \\
\hline  
\end{tabular}
}
\label{tab:stations}
\tablefoot{Due to operational issues, data from the Badary (BD), Ulsan (KU), and Yonsei (KY) stations were excluded from the analysis.}
\end{table}

Image reconstruction was performed using the RML method, 
implemented in the \texttt{eht-imaging} software package \texttt{ehtim} \citep{Chael_2016, 2018ApJ...857...23C}. Unlike traditional techniques, the RML approach avoids using the inverse Fourier transform of the visibilities, $V$, during the imaging process. Instead, it integrates forward modeling with regularization to derive an image, $I$, that minimizes the objective function:

\begin{equation}
    J(I) = \sum_{\rm{data\ terms}}\alpha_D\chi^2_D(I, V) - \sum_{\rm{reg.\ terms}}\beta_R S_R(I),
    \label{eq:of}
\end{equation}The first term of this equation promotes consistency with the observed data, 
while the second term enforces regularization to ensure a physically meaningful image. 
The relative contributions of these terms are controlled by the hyperparameters $\alpha_D$ and $\beta_R$, which balance data fidelity against regularization. Using this method, we achieve higher angular resolution and image fidelity compared to traditional techniques 
like \texttt{CLEAN} \citep[e.g.,][]{eht_2019,eht_2022}.

The imaging 22\,GHz RadioAstron data obtained on April 25, 2016, was carried out in several steps, beginning with the reconstruction of the ground array. We initialized the process with a Gaussian prior image (full width at half maximum = 150\,$\mu$as), refining the solution iteratively. This was done across a 2\,mas field of view, discretized onto a $300\times300$ pixel grid. We incorporated closure phases and logarithmic closure amplitudes, accounting for non-closing systematic errors of 1\% of visibility amplitude. 
After achieving convergence, we performed self-calibration to derive gain solutions and improve image fidelity.In subsequent iterations, we expanded our approach. We first included visibility phases and amplitudes, followed by a comprehensive integration of both closure quantities and complex visibilities. The final stage involved applying this strategy to include the SRT, using the ground array image as a prior.

\begin{figure*}
\centering
\includegraphics[width=\textwidth]{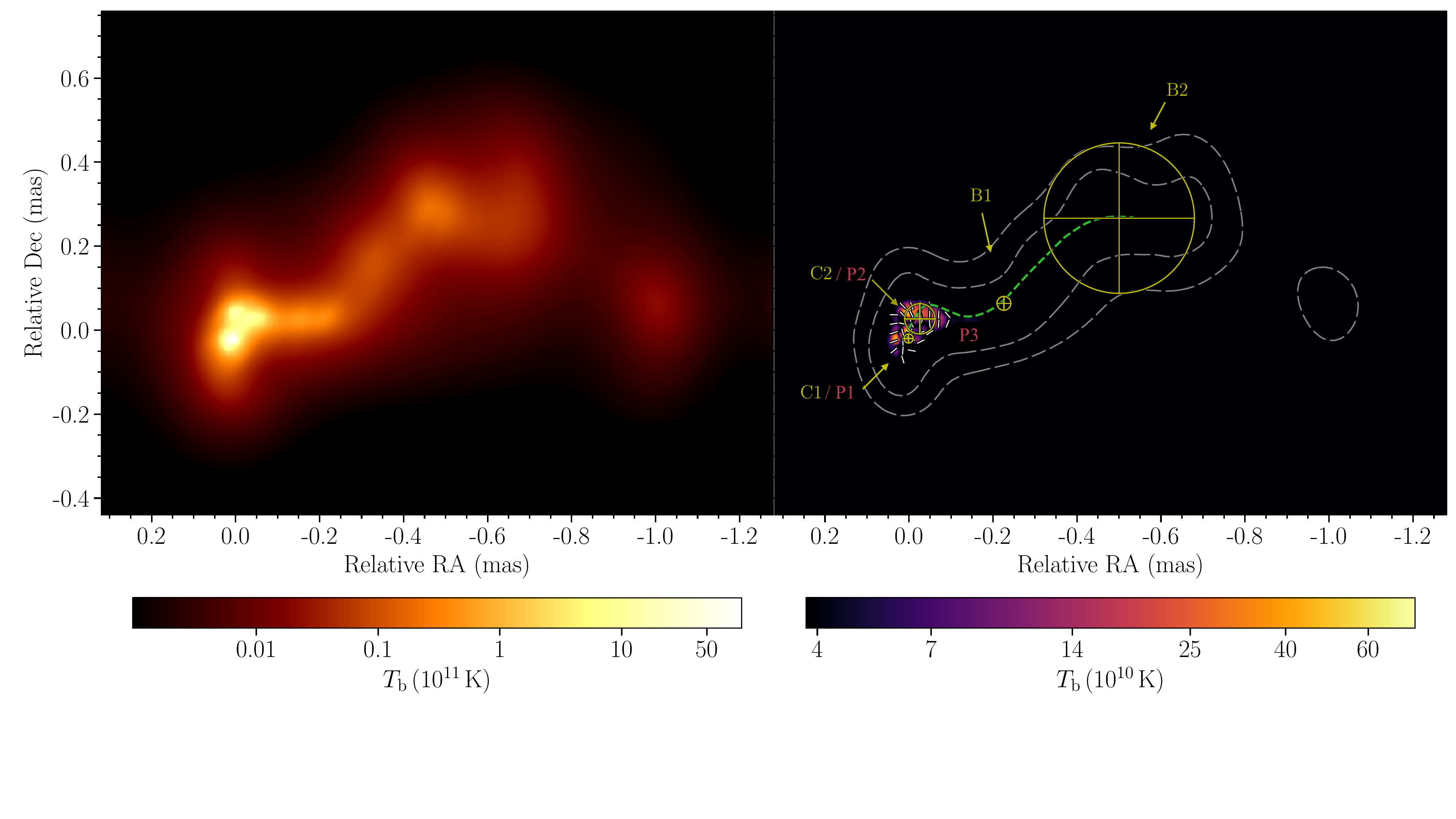}
\caption{22\,GHz image of \object{OJ\,287} on April 25, 2016. Left: Total intensity image shown in brightness temperature ($T_b$) units, calculated using the relation $T_b = I_\nu c^2 / (2k\nu^2)$, where $I_\nu$ is the specific intensity per pixel. The pixel size used is 5\,$\mu$as, derived from the total field of view and the image resolution (320 pixels across). The peak $T_b$ reaches $\sim6 \times 10^{12}$\,K in the core region. Right: Total intensity contours (dashed white lines) spaced logarithmically between 0.01\% and 0.1\% of the peak intensity. The color map shows the magnitude of the linearly polarized flux. Regions where polarization is not reliably detected downstream have been masked. White ticks indicate the EVPA orientation. Yellow circles mark the centroids of the two-dimensional Gaussian-fitted components. Polarized features P1, P2, and P3 are shown in purple. The central dashed green curve traces the jet ridgeline, as derived from the smoothed image (see Sect.~\ref{sec:analysis_results}).}
     \label{fig:main}
\end{figure*}

To determine optimal relative weights for the regularization terms, we systematically varied their values across a search grid and selected the combination yielding the highest image fidelity and best data fit. This involved testing various values for relative entropy, total variation (tv), total squared variation (tv2), total variation $\ell$2 with logarithmic regularizer (tv2log), and the $\ell$1 norm. The tv2log regularizer combines the tv norm with an $\ell$2 norm and incorporates a logarithmic transformation, 
balancing between promoting sparsity and smoothness while handling large dynamic ranges in image intensities \citep[see also][]{2023A&A...676A.114S}.
The final combination of values were: relative entropy: 0.1, tv: 0, tv2: 0.01, tv2log: 0.05, and $\ell$1: 1, resulting $\chi^2_\mathrm{phas}$ : 1.04, and $\chi^2_\mathrm{logamp}$: 1.98. Details of the parameter selection process are provided in Appendix \ref{appA}. 
We applied the same imaging procedure to the 43\,GHz data, 
with the resulting images shown in Figs.~\ref{fig:main} and \ref{fig:bu}.

We imported the fully calibrated RadioAstron data into \texttt{Difmap} \citep{1997ASPC..125...77S}, an interactive program for synthesis imaging. Using the \texttt{MODELFIT} algorithm, we parameterized the jet brightness distribution 
with two-dimensional circular Gaussian components. The uncertainties for each component parameter were formally assessed based on the local S/N 
in the surrounding image \citep{1999ASPC..180..301F,2005astro.ph..3225L,2012A&A...537A..70S}. For the flux density measurements, we adopted a more conservative 20\% uncertainty, to account for the increased calibration uncertainties inherent to space VLBI observations at 22\,GHz. This choice is more appropriate than the 10\% typically used in ground-based VLBI at lower frequencies \citep{2009AJ....138.1874L}, given the limited system temperature  and gain information available for some antennas and the lower S/N on the space baselines.
All parameters for the fitted Gaussian components are listed in \autoref{table:knots}.

\begin{table}[h!]
\centering
\caption{Model-fitting results using circular Gaussian components with \texttt{Difmap} for the 22\,GHz RadioAstron data of \object{OJ\,287} from April 25, 2016.}
{\small
\begin{tabular}{ccccc}
\hline
\hline
ID & S & r & $\theta$ & FWHM \\
 & (Jy) & (mas) & ($^\circ$) & (mas) \\
\hline
C1 & 1.52 $\pm$  0.30 & - & -  & 0.02 $\pm$ 0.002 \\
C2 & 1.84 $\pm$ 0.40 & 0.05 $\pm$ 0.01 & $-$29.9 $\pm$ 2.0 & 0.07 $\pm$ 0.008  \\
B1 & 0.12 $\pm$ 0.02 & 0.24 $\pm$ 0.02 & $-$79.7 $\pm$ 3.0 & 0.03 $\pm$ 0.005 \\
B2 & 0.34 $\pm$ 0.06 & 0.58 $\pm$ 0.03 & $-$61.2 $\pm$ 5.0  & 0.36 $\pm$ 0.060  \\
\hline
\end{tabular}
}
\label{table:knots}
\end{table}

\subsection{Imaging and model-fitting: Polarization}

Following total intensity imaging and polarimetric calibration, we reconstructed polarized images using \texttt{ehtim}. 
We minimized the objective function in Eq.~\ref{eq:of}, incorporating both polarimetric visibilities ($\tilde{P} = \tilde{Q} + i\tilde{U}$) and 
the visibility-domain polarization ratio ($\tilde{m} = \tilde{P} / \tilde{I}$), the latter being immune to residual station gain errors from Stokes $I$ imaging. 
For image reconstruction, we employed two regularizers: (1) the Holdaway-Wardle regularizer \citep[hw;][]{Holdaway1990} to constrain pixel polarization fraction below 0.75 
(the theoretical maximum for synchrotron radiation), and (2) the polarimetric total variation (ptv) regularizer \citep{Rudin1992} to ensure smooth polarization transitions. 
We reconstructed total intensity and linearly polarized intensity images independently, solving for the fractional polarization ($m$) and the electric vector position angle (EVPA; $\chi$) 
in each pixel. The iterative gradient descent process used the previous iteration's output blurred by a 20\,$\mu$as Gaussian kernel, 
with final values of hw:1 and ptv:1 chosen for optimal data fit and image fidelity. 
The pipeline alternated between polarimetric objective function minimization and D-term calibration, maximizing consistency between self-calibrated data and corrupted image reconstructions. After D-term solutions were obtained (see Appendix \ref{appA}), we blurred the polarimetric image and repeated the imaging-calibration cycle until convergence.

In addition to correcting for instrumental polarization, VLBI polarimetric analyses require accurate absolute EVPA calibration. To this end, we applied a correction based on a close-in-time VLBA observation at 43\,GHz from the BEAM-ME project (successor to the VLBA-BU-BLAZAR program), which reported an EVPA of $-12^\circ \pm 2^\circ$ for OJ,287 on April 22, 2016. This value was used to calibrate the RadioAstron EVPA measurements. As an independent reference, we also considered a single-dish Effelsberg observation of OJ,287 at 10.45\,GHz, conducted on April 12, 2016, which measured $\text{EVPA} = 3^\circ.36 \pm 0.75$ \citep{2018A&A...619A..88M}, obtained within the framework of the MOMO monitoring program \citep{2015ATel.8411....1K,Komossa2023_MOMO}. Finally, we computed the net polarization, $|m|_{\rm net}$, which reflects the degree of alignment in polarization direction across the source, and the average linear polarization fraction, $\langle |m| \rangle$, which quantifies the typical polarization strength, independent of the directional coherence following \cite{Akiyama_2021}:

\begin{equation}
  |m|_{\rm net} = \sqrt{\left( \sum_i Q_i \right)^2 + \left( \sum_i U_i \right)^2 } / \sum_i I_i
  \label{eq:mnet}
\end{equation}

\begin{equation}
 \langle |m| \rangle = \sum_i \sqrt{Q_i^2 + U_i^2}/ \sum_i I_i 
 \label{eq:mavg}
.\end{equation}These equations yield 0.45\% and 5.5\%, respectively. Although several bright calibrator sources (0716$+$714, 3C\,345, 0059$+$581, 4C\,+38.41, and 4C\,+45.51) were observed as part of the experiment to serve as fringe finders and potential D-term calibrators, we ultimately derived the instrumental polarization calibration using OJ\,287 itself, which provided more reliable solutions due to its strong, compact polarized structure and longer integration time.

\section{Results}
\label{sec:analysis_results}

\begin{figure*}
\centering
\includegraphics[width=0.245\textwidth]{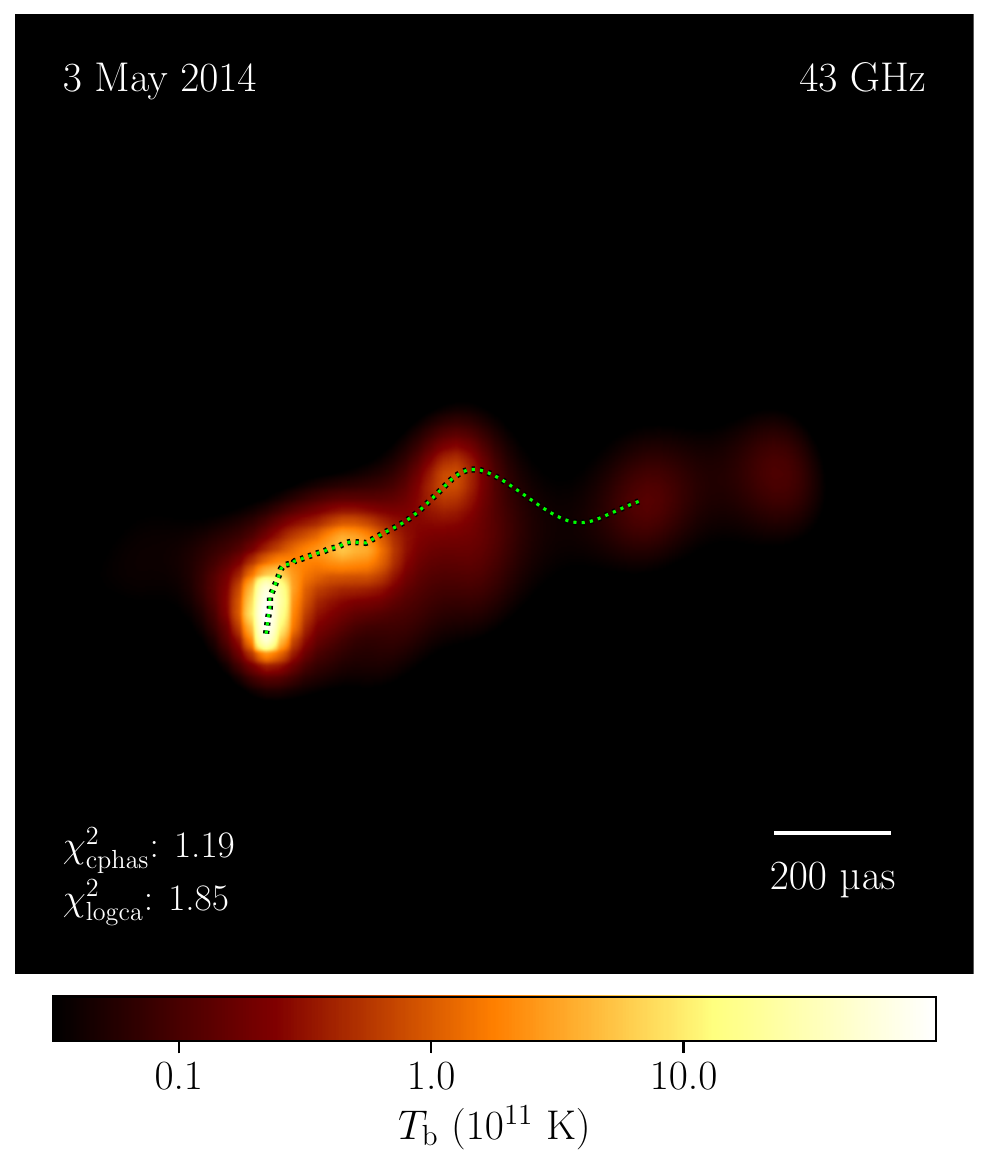}
\includegraphics[width=0.245\textwidth]{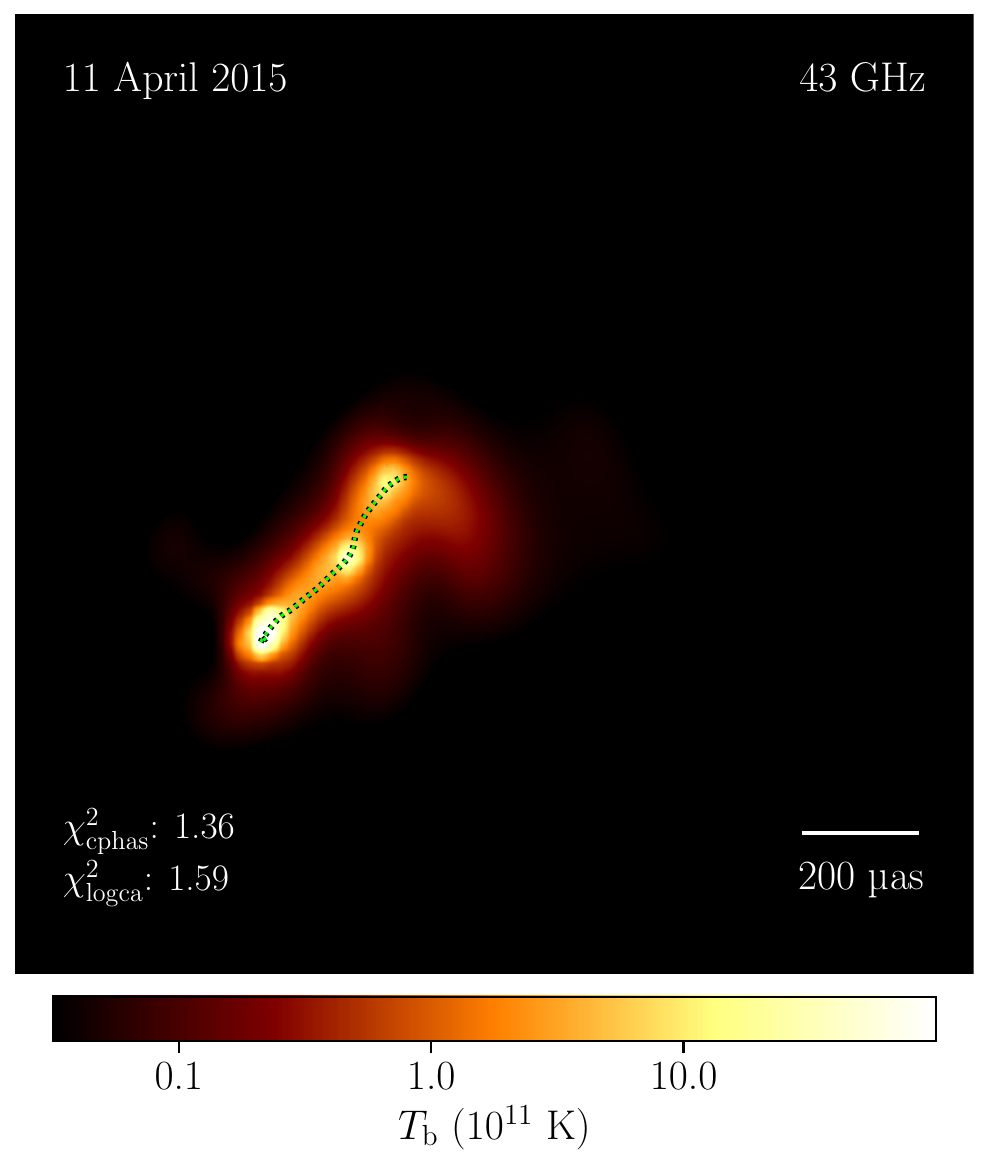}
\includegraphics[width=0.245\textwidth]{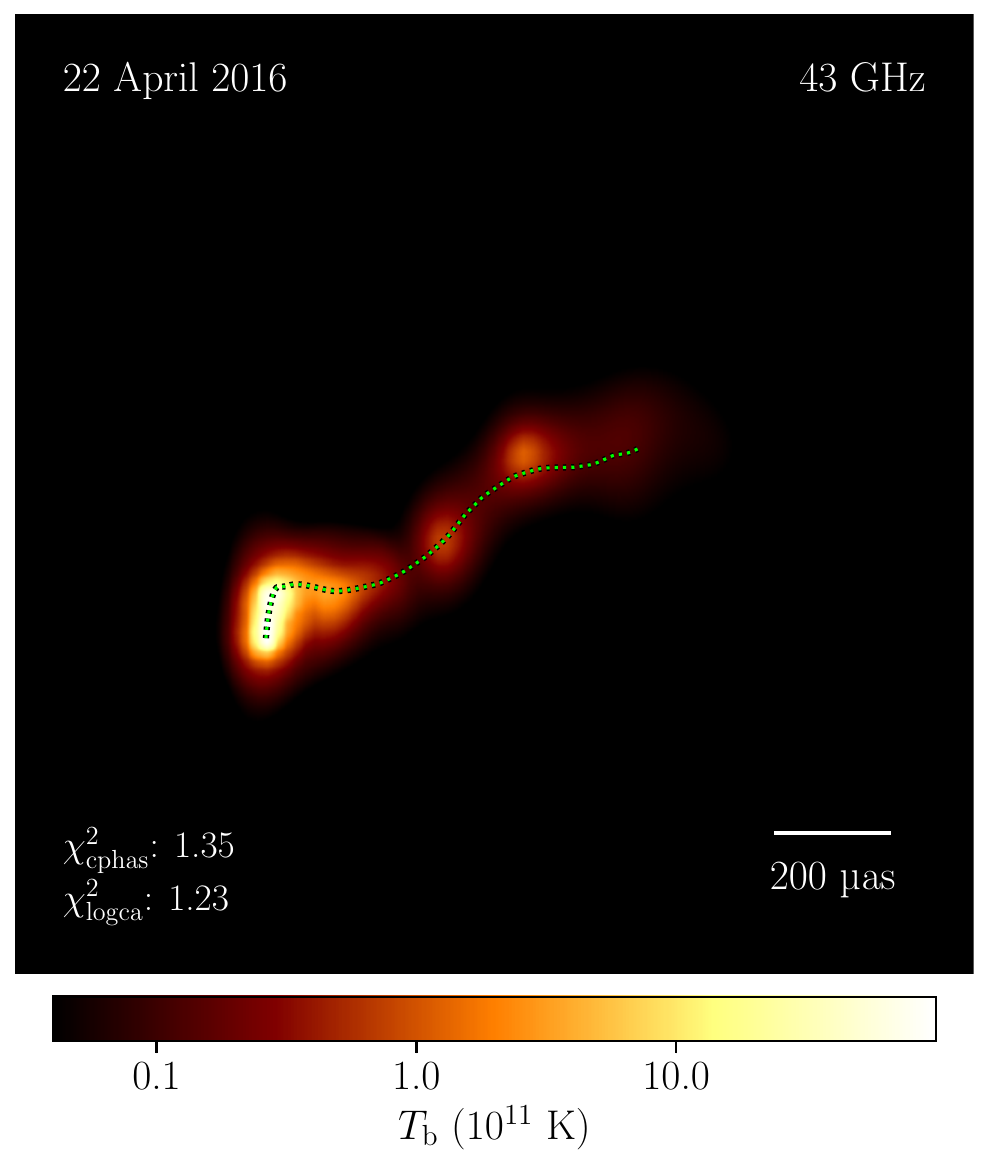}
\includegraphics[width=0.245\textwidth]{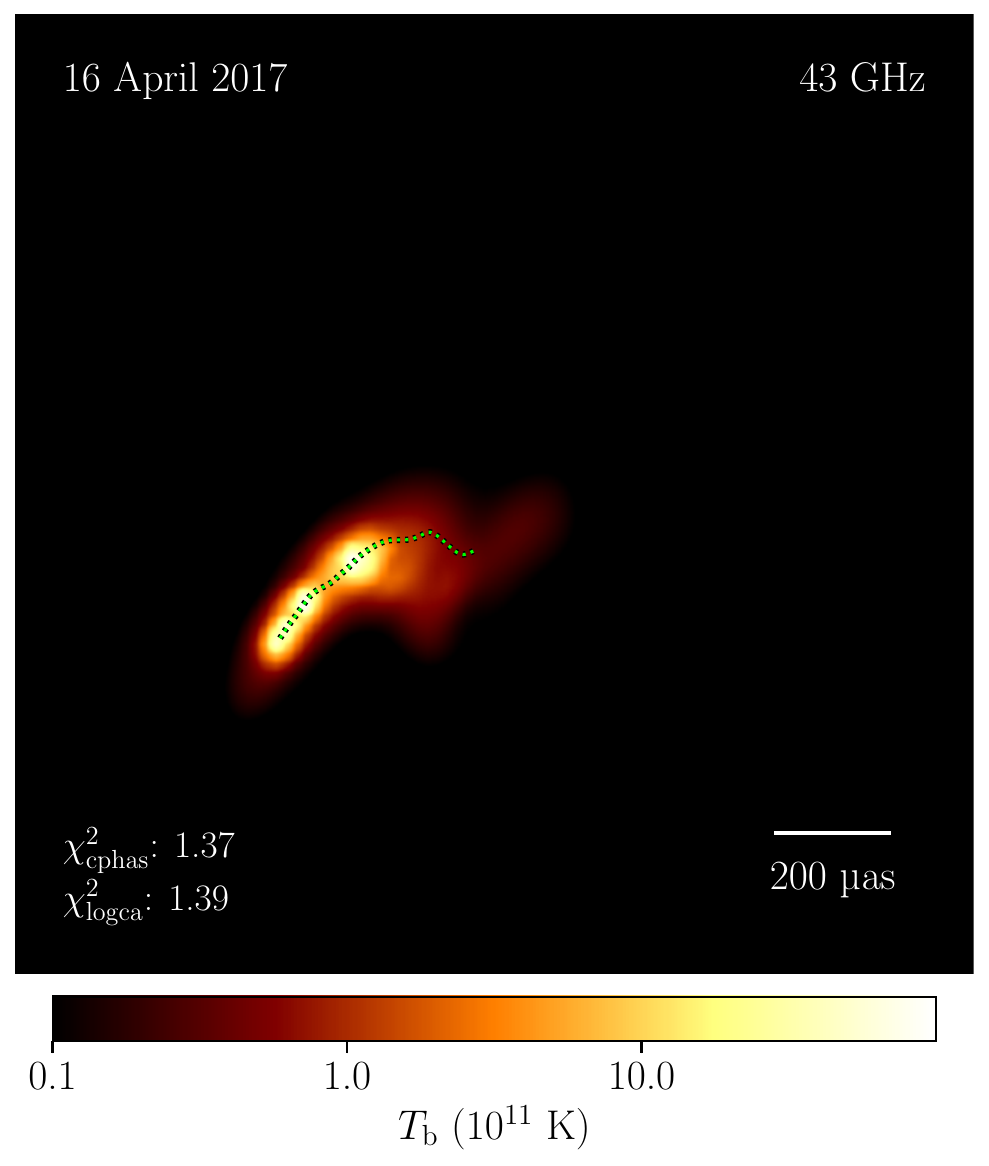}
  \caption{Multi-epoch 43\,GHz VLBI images of \object{OJ\,287} observed on May 3, 2014, April 11, 2015, April 22, 2016, and April 16, 2017. 
Each panel shows the source morphology in total intensity as reconstructed by \texttt{ehtim}. 
The color scale is the same for all days and is in units of brightness temperature; an angular scale indicator is included in the bottom-right corner of each image (200\,$\mu$as) for reference. 
The $\chi^2$ statistics used as diagnostics of the model fit quality  --- the closure phase reduced $\chi^2$ ($\chi^2_\mathrm{cphase}$) and  the logarithmic closure amplitude reduced $\chi^2$ ($\chi^2_\mathrm{logamp}$) --- 
are displayed in the bottom-left corner of each image. 
The ridgelines, delineating the jet's axis for each epoch, are overlaid as dashed lime lines, with a black outline for enhanced visibility. 
In this image sequence we can clearly see the jet's remarkable PA evolution over the four years.}
     \label{fig:bu}
\end{figure*}

\subsection{Jet morphology}

The left panel of Fig.~\ref{fig:main} shows the total intensity image of \object{OJ\,287} from our RadioAstron observations. 
The core region exhibits a remarkable and atypical structure, deviating from the standard core-dominated morphology typically observed in blazars. 
The jet emanating from the core initially propagates northwest before undergoing a sequence of dramatic directional changes. 
The first significant bend occurs at $\sim$100$\,\mu$as from the core, where the jet trajectory sharply changes direction. 
Subsequently, two additional bends are observed at $\sim$300\,$\mu$as and $\sim$550\,$\mu$as to the west, forming a ribbon-like morphology. 
This complex trajectory may reflect a possible interaction with the surrounding medium, precession of the jet axis, inherent instabilities within the jet flow, 
or all mechanisms working in combination.

\indent Our model-fitting analysis reveals four distinct Gaussian components. 
The core region is resolved into: C1, with a flux density of 1.5\,Jy, 
located upstream and likely representing the VLBI core, and C2, appearing as the brightest knot with a flux density of 1.8\,Jy. 
We note that compared to our 2014 observations where C1 and C2 showed flux densities of 0.5\,Jy and 1.17\,Jy, respectively \citep{2022ApJ...924..122G}, 
the current enhanced brightness of C1 suggests a recent ejection event where a new component is becoming optically thin at 22\,GHz. 
Further downstream, components B1 and B2 are located at the bending points, marking the transition from the compact core region to the broader jet structure.
We computed the brightness temperatures for each component using the relation 
T$_\mathrm{b} = 1.22 \times 10^{12} \left( S_\nu / \left( \theta_\mathrm{obs}^{2} \nu^{2} \right) \right) \left( 1+z \right)$ (K) \citep[e.g.,][]{2012A&A...544A..34P}, where $S_{\nu}$ is the component flux density in Jy, $\theta_{\rm obs}$ is the component full width at half maximum in mas, and $\nu$ is the observing frequency in GHz, and $z$ is the redshift. The values decrease systematically with distance from the core down the jet: 
$T_\mathrm{b,C1} = (1.3 \pm 0.28) \times 10^{13}\,\mathrm{K}$, 
$T_\mathrm{b,C2} = (1.2 \pm 0.31) \times 10^{12}\,\mathrm{K}$, 
$T_\mathrm{b,B1} = (4.4 \pm 1.5) \times 10^{11}\,\mathrm{K}$, and 
$T_\mathrm{b,B2} = (8.6 \pm 3.0) \times 10^{9} \, \mathrm{K}$.\\
\\
\indent The exceptionally high $T_\mathrm{b}$ of C1 exceeds both the inverse Compton limit ($\sim10^{12}$ K; \citealt{1969ApJ...155L..71K}) 
and the equipartition brightness temperature ($\sim10^{11}$ K) limits \citep{1994ApJ...426...51R}, 
requiring Doppler factors $\delta_\mathrm{j}\sim10\text{--}30$ for reconciliation. 
This agrees with estimates from VLBA-BU-BLAZAR monitoring at 43\,GHz ($\delta_\mathrm{j}=8.6\pm2.8$; \citealt{2022ApJS..260...12W}), 
suggesting a jet viewing angle $\theta_\mathrm{j}\approx3^\circ\text{--}8^\circ$. 
Such extreme brightness temperatures could indicate strong Doppler boosting, 
departure from equipartition conditions, or potentially more exotic emission mechanisms such as relativistic proton emission \citep{Kardashev2000} 
or coherent emission processes \citep{1998MNRAS.301..414B}. 
See also the results and discussion in \citet{2016ApJ...820L...9K,2020AdSpR..65..705K}.
Meanwhile, the decreasing $T_\mathrm{b}$ trend along B1 and B2 reflects the expected jet expansion and energy dissipation with distance from the core \citep[e.g.,][]{Roeder2025}.

Our polarimetric analysis revealed a prominent structure in the inner jet (extending within the first $\sim100\,\mu\text{as}$), 
as shown in the right panel of Fig.~\ref{fig:main}. In this region, the EVPAs are perpendicular to the jet axis, 
highlighting the dominance of the poloidal magnetic field component, 
favoring the development of kink instabilities \citep[e.g.,][]{2004AIPC..703..308N}. We describe the polarization structure in terms of three distinct features, 
namely P1, P2, and P3. P1, the innermost component, exhibits a polarized intensity of $P_1 = 0.07 \pm 0.01$\,Jy,
a fractional polarization of $m_1 = 3.0\% \pm 0.3\%$, and an EVPA of $\chi_1 = 42^\circ \pm 5.0^\circ$.
P2, located farther downstream, exhibits $P_2 = 0.07 \pm 0.01$\,Jy, $m_2 = 18.5\% \pm 2.0\%$, and $\chi_2 = 48^\circ \pm 5.0^\circ$. The distinction between P1 and P2 was made based on their proximity, morphology, and the presence of a gradient in both polarized intensity and total intensity. 
While there is a continuity in emission between P1 and P2, P1 is identified as the brightest feature in the southernmost region, whereas P2 corresponds to a distinct intensity peak farther downstream. P3, the outermost feature, displays the lowest polarized intensity ($P_3 = 0.05 \pm 0.01$,Jy), 
but maintains significant fractional polarization ($m_3 = 20.5\% \pm 3.3\%$), with an EVPA of $\chi_3 = 38.8^\circ \pm 5.0^\circ$. 
The elevated fractional polarization values in P2 and P3 indicate that, despite their lower polarized intensities, 
the magnetic field retains a significant degree of alignment, which could arise from the effects of shocks or compression in the jet flow. 
All uncertainties in $P$, $m$, and $\chi$ were calculated following \citet{2022ApJ...924..122G}.

\subsection{Jet ridgeline analysis}

To quantify the jet structure and PA of \object{OJ\,287}, we performed a ridgeline analysis as described in \cite{2022ApJ...932...72Z}. 
First, we generated a smoothed version of the reconstructed image by applying a Gaussian blur with a width equivalent to twice the image resolution (47\,$\mu$as). 
The smoothed images were transformed into polar coordinates, centered on the jet origin. To construct the ridgeline, the polar-transformed images were sliced radially, 
and transverse intensity profiles were extracted along the jet axis. For each slice, we fitted a Gaussian function to the intensity profile to identify the peak flux density position, 
which was subsequently transformed back into Cartesian coordinates. Finally, the resulting set of peak positions was interpolated using a cubic spline to obtain a continuous ridgeline tracing the jet path. All uncertainties were estimated using the covariance matrix from Gaussian fits to derive the uncertainty of the peak position ($\sigma_{x_0}$). 
The radial uncertainty ($\sigma_r$) was set equal to $\sigma_{x_0}$, while the angular uncertainty ($\sigma_\theta$) was approximated as $1/r$, 
accounting for the decreasing angular precision with increasing radial distance. Failed Gaussian fits were excluded from the analysis. Using standard error propagation techniques, 
the ridgeline positions and their associated uncertainties were transformed back into Cartesian coordinates. In addition to the ridgelines derived from the data analyzed in this work, 
we also included ridgeline data for \object{OJ\,287} from four selected epochs at 43\,GHz, obtained from the publicly available VLBA-BU-BLAZAR database. 
These epochs span April 2014 to 2017 (May 3, 2014, April 11, 2015, April 22, 2016, and April 16, 2017). 
Their comparison allowed us to investigate annual changes in jet orientation and check their agreement with theoretical predictions (see the discussion in Sect.~\ref{sec:bbh}). \\

As can be seen in Fig.~\ref{fig:bu}, the jet demonstrates a clear temporal evolution. 
In 2014, \object{OJ\,287} appears smooth and extended, displaying modest curvature and reaching $\sim$600\,$\mu$as apparent length from the core. 
The 2015 epoch reveals a shorter jet configuration, with emission confined to $\sim$300\,$\mu$as. By 2016, the jet regains its extended structure to $\sim$600\,$\mu$as, 
showing also strong morphological agreement with 22\,GHz RadioAstron image, reinforcing the jet structure across frequencies. 
The 2017 epoch exhibits again a compact, bright inner region with minimal downstream emission, similar to the confined structure observed in 2015. 
These variations in apparent jet extension and morphology can result from multiple factors, including changes in Doppler boosting or jet precession due to instabilities, Lense-Thirring effects, or the presence of an SMBHB system. Additionally, differences in data quality, such as variations in dynamic range or noise level, can affect the detectability of extended jet features and may also contribute to the observed morphological differences.

\section{Discussion}
\label{sec:discussion}

\subsection{OJ\,287 multi-scale jet bending}

The revelation of a ribbon-like morphology with three distinct bends within 650\,$\mu$as in the jet of \object{OJ\,287} revealed by RadioAstron observations is a unique occurrence. 
Unlike previous high-resolution VLBI studies of \object{OJ\,287} \citep[e.g.,][]{2017ApJ...838...78H, 2022A&A...658L..10L, 2022ApJ...924..122G, 2022ApJ...932...72Z}, 
which could only resolve the first bend, our RadioAstron observations from April 25, 2016, have provided the first direct imaging of the full multi-scale jet reorientation 
within the acceleration and collimation zone. According to \citet{2008Natur.452..966M,Homan_2015}, the acceleration and collimation zone in blazars typically extends 
to $\sim10^4$--$10^6$ Schwarzschild radii ($R_S$) from the central black hole. For \object{OJ\,287}, using the latest mass estimate for the central supermassive black hole
of $\sim10^8$ $M_{\odot}$ \citep{Komossa2023_Impact, Komossa2023_MOMO} and a luminosity distance of 1.6\,Gpc, 
this corresponds to an angular scale of approximately 6.1--610 $\mu$as on the sky. 
The resolution of $\sim$47~$\mu$as achieved in this work (improved to $\sim$15\,$\mu$as after \texttt{ehtim} imaging), offered us the ability to capture these multiple bends and reveal that the jet undergoes 
a substantially more complex evolution than previously assumed.

\subsection{Jet position angle variations and SMBHB precession predictions } 
\label{sec:bbh}

\begin{figure}[h!]
\centering
\includegraphics[width=\columnwidth]{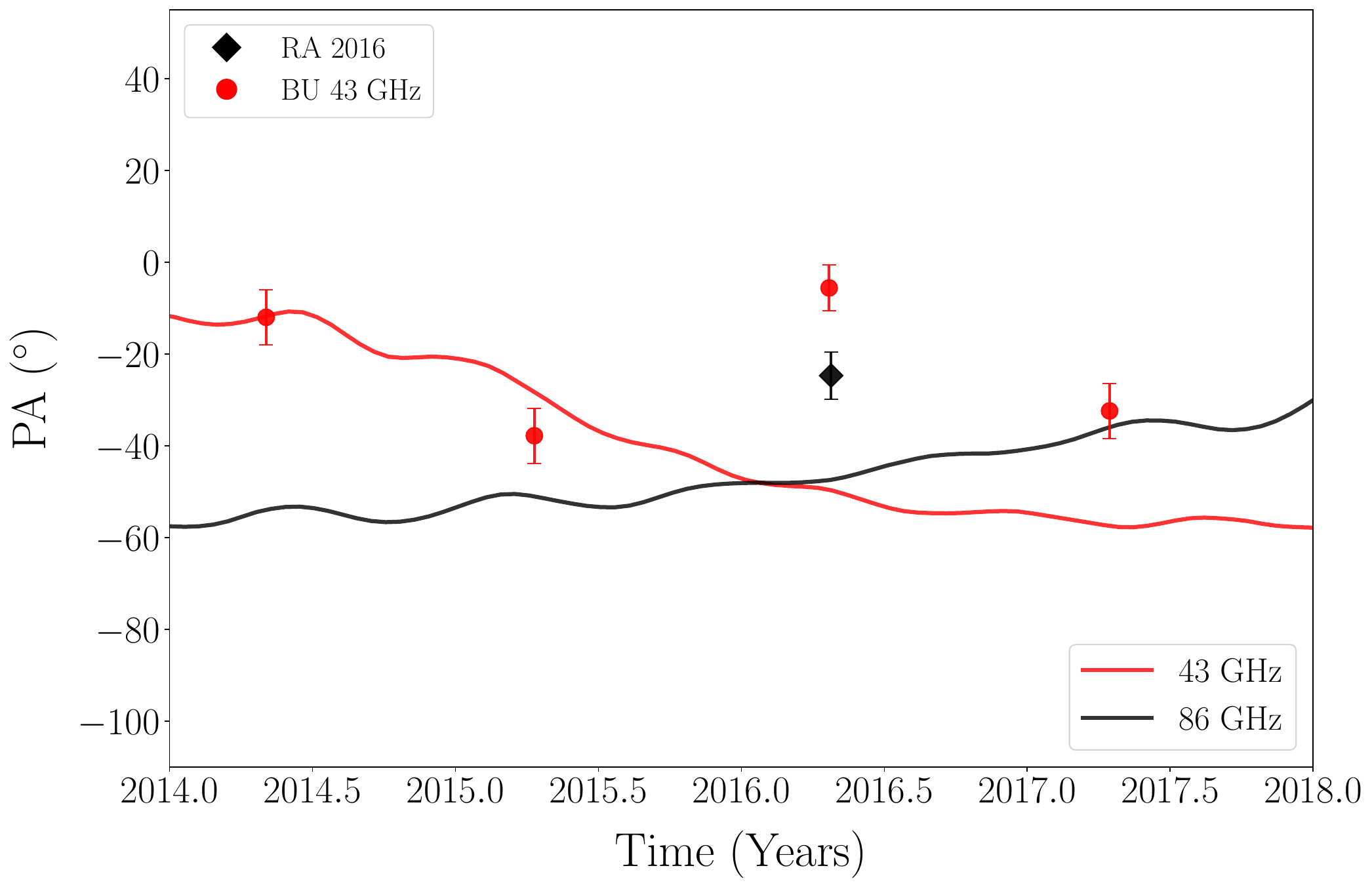}
  \caption{Fits to the variations in the PA of the radio jet at 43\,GHz and 86\,GHz based on the expected jet precession in the disk model by \cite{Dey2018}. The red and black lines correspond to the model predictions at 43\,GHz and 86\,GHz, respectively. The innermost PA value from RadioAstron is marked with black diamond, and VLBA-BU-BLAZAR data points are marked with red circles.} 
     \label{fig:bbh_model}
\end{figure}

Position angle variations in the \object{OJ\,287}'s jet can provide crucial insights into properties of the binary black hole in this system. 
\cite{Dey2018} conducted a Bayesian analysis to explore how the PA variations observed in the source at multiple frequencies (15, 43, and 86\,GHz) 
could be explained by precession mechanisms linked to the SMBHB system dynamics. Their study explored two potential origins for the precession: 
the evolution of the primary black hole's spin, which attributes the variations to Lense-Thirring precession caused by the primary black hole's spin (spin model), 
and the precession of the angular momentum of the inner accretion disk, which links to the hydrodynamical evolution of the inner disk's angular momentum under 
the gravitational influence of the secondary black hole (disk model). While both models describe the data well, the accretion disk precession model provided 
a more consistent description of PA variations observed at higher frequencies (e.g., 86\,GHz in \citealt{Dey2018}), making it particularly relevant for comparison with our data here.

\begin{figure}[ht!]
\centering
\includegraphics[width=\columnwidth]{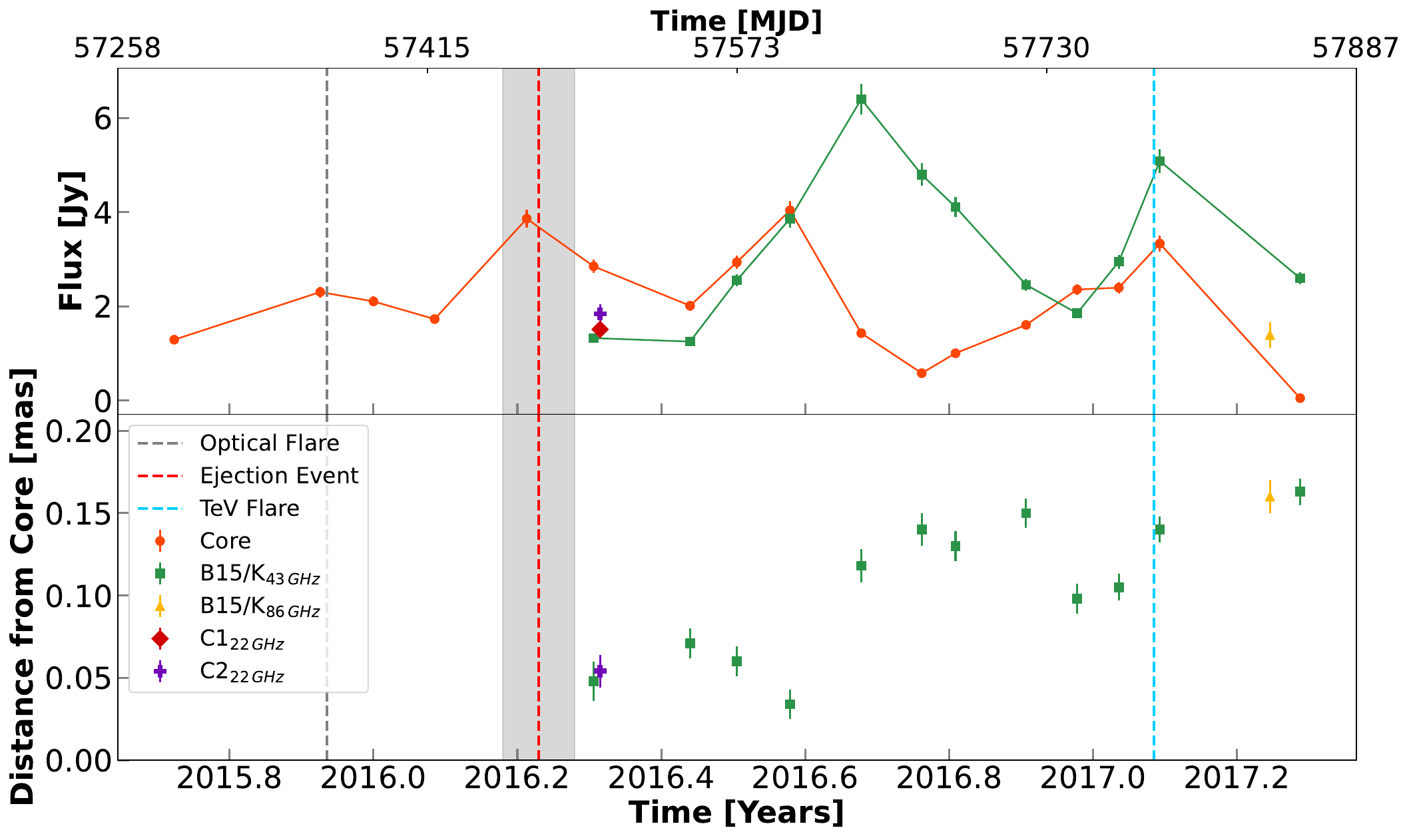}
\caption{Flux density and radial core separation evolution of jet components within 0.2 mas of the core in \object{OJ\,287} at 22, 43, and 86\,GHz from September 2015 to April 2017. Top panel: Flux density evolution over time, showing the core (red circles), a knot identified as B15 at 43\,GHz and K at 86\,GHz (green square and yellow triangle, respectively), and components C1 (magenta diamond) and C2 (purple cross) identified in this work at 22\,GHz. Error bars indicate measurement uncertainties. Bottom panel: Radial core separation projected distances. A TeV flare (dashed cyan vertical line) and the centenary optical flare (dashed gray vertical line) are marked. The dashed red vertical line and the gray shaded region show the estimated ejection time and its uncertainty for knot B15 or K \citep{2022ApJS..260...12W}.}
\label{fig:kinem_bu}
\end{figure}

To test these theoretical predictions, we analyzed jet PAs using both our RadioAstron observations and the VLBA-BU-BLAZAR data at 43\,GHz. We derived the mean PA for each epoch by averaging ridgeline points within the first bright jet component from the core, consistent with the model resolution limit and the region of the jet consistently visible in all epochs. To establish robust uncertainty estimates, we implemented a statistical bootstrapping method. This involved randomly resampling with replacement from our original PA measurements 10,000 times, creating 10,000 simulated datasets. For each of these simulated datasets, we calculated the circular standard deviation, producing a distribution of possible standard deviations. This approach provides a reliable error estimation that accounts for the circular nature of angle measurements (where 0$^{\circ}$ and 360$^{\circ}$ are identical), while reducing potential biases from outliers or epochs with limited sampling. For the 22\,GHz 2016 data, the presence of RadioAstron ensured angular resolutions comparable to the 86\,GHz observations.

Our analysis yielded an innermost PA ($\leq$100\,$\mu$as translate to $\sim$0.45\,pc projected distance)of $-25^\circ \pm 5^\circ$ for the RadioAstron 2016 observations. For the VLBA-BU-BLAZAR 43\,GHz data, we obtained values of $-12^\circ \pm 6^\circ$ on May 3, 2014, $-38^\circ \pm 6^\circ$ on April 11, 2015, $-6^\circ \pm 5^\circ$ on April 22, 2016, and $-32^\circ \pm 6^\circ$ on April 16, 2017. These measurements reveal a clear evolution of the jet PA within $\sim$100\,$\mu$as, with a change of $\sim25^\circ$ in the 43\,GHz measurements across the four epochs. This change represents a significant reorientation of the jet direction. In the context of the disk model, the PA variations observed in our VLBI datasets align well with the predictions only for the first 2 epochs, with the 2016 data (at both 43 and 86 GHz) deviating from the models, while the 2017 data follow the 86 GHz prediction as shown in Fig. \ref{fig:bbh_model}. 

\subsection{Relation with multiwavelength flares and jet dynamics}

A major optical outburst occurred in \object{OJ\,287} on December 5, 2015 (MJD 57363, corresponding to 2015.93), with the source reaching its brightest optical level in three decades. This event is also known as the "Centenary Flare," as it was predicted with astonishing accuracy by the SMBHB model \citep{Valtonen2006}. The model suggested that the outbursts are attributed to the secondary black hole periodically impacting the accretion disk of the primary supermassive black hole, producing thermal bremsstrahlung radiation visible as bright optical flares \citep{LehtoValtonen1996, ivanov1998}. It is expected that such interactions in binary systems can enhance jet activity through multiple mechanisms: disk perturbations that increase accretion rates \citep{LehtoValtonen1996, Sundelius1997, angelo2016}, magnetic field intensification associated with the orbital motion of the secondary \citep[e.g.,][]{Palenzuela2010, Gold2014}, and direct gravitational torques capable of influencing the jet-launching region \citep[e.g.,][]{Hayasaki2008, DOrazio2013}.

Subsequently, on March 15, 2016 (2016.21), \object{OJ\,287} exhibited another strong flare detected in both the optical and radio bands, originating from the jet. The optical flare, comparable in strength to the December 2015 outburst, displayed polarization characteristics indicative of synchrotron emission, suggesting a jet-driven origin rather than a direct disk impact \citep{Gupta2017, Kushwaha2018}. Simultaneously, a significant radio flare was observed, likely triggered by a propagating shock wave within the relativistic jet \citep{Gupta2017}. Kinematic analysis of VLBA-BU-BLAZAR data \citep{2022ApJS..260...12W} confirmed the ejection of component B15 at $2016.23 \pm 0.10$, characterized by a proper motion of $0.28 \pm 0.02$\,mas/year ($5.38 \pm 0.32$\,c). One month before this ejection, the core experienced a prominent flare where the flux density increased from $1.7$\,Jy to $3.9$\,Jy (130\%), consistent with the typical behavior observed when a new component traverses the radio core \citep{2008ASPC..386..437M}. 

An independent study by \cite{2022A&A...658L..10L} 
at 86\,GHz detected the same component, designating it as K, with a measured propagation speed of $\sim$0.32\,mas/year ($\sim$4.8\,c), yielding an ejection time of $\sim$2016.3. Their analysis suggests that the TeV flaring activity observed February 1-4, 2017, was triggered when this new jet feature passed through a recollimation shock S1 at $\sim$0.1\,mas from the radio core, a stationary jet feature mentioned in multiple previous studies \citep{2017A&A...597A..80H,2022ApJ...924..122G}. Figure~\ref{fig:kinem_bu} illustrates the emergence and propagation of this new component, tracking both its motion away from the core and its flux density evolution over time, while also showing the temporal relationship between optical, radio and TeV flaring events.

\subsection{Comparison with single-dish radio polarization monitoring}

\citet{2018A&A...619A..88M} demonstrated that \object{OJ\,287} underwent a prolonged clockwise rotation of its radio EVPA during 2016, based on single-dish Effelsberg observations carried out within the framework of the MOMO monitoring program \citep{2015ATel.8411....1K,Komossa2023_MOMO}. Their analysis revealed that this rotation occurred within the jet core as observed by the VLBA-BU-BLAZAR monitoring program at 43\,GHz. This region corresponds to a distance of approximately $\sim10^4$--$10^6$$R_S$ from the central black hole (assuming a mass of $10^8$--$10^{10}$~$M_{\odot}$), placing it in the outer acceleration and collimation zone. Our RadioAstron data now allow us to examine this region with enhanced resolution.

The radio EVPA evolution in 2016 reported in \cite{2018A&A...619A..88M} was attributed to the possible bending of the inner jet within the 43\,GHz core, which has a projected size of about 0.15--0.2~mas. Our RadioAstron imaging results of the same year (Fig.~\ref{fig:main}) show that there is indeed strong bending of the inner jet at those spacial scales.

A noteworthy detail emerges when comparing the EVPA rotation rate in 2016 reported in Fig.~4 of \cite{2018A&A...619A..88M} with the inner jet morphology of the same year as revealed by RadioAstron. The EVPA rotation rate decreased from 4$^\circ$/day to 1$^\circ$/day shortly after the onset of the rotation event, as seen only in the 10.45~GHz data of that figure.\footnote{Radio polarization data at lower frequencies, being affected by a dominant stable polarization component that was attributed to the large-scale jet contribution, miss the start of the rotation event.} This deceleration can be attributed to a more tightly wound helical structure near the jet origin, precisely what our 2016 RadioAstron image reveals.

\begin{figure}
\includegraphics[width=\columnwidth]{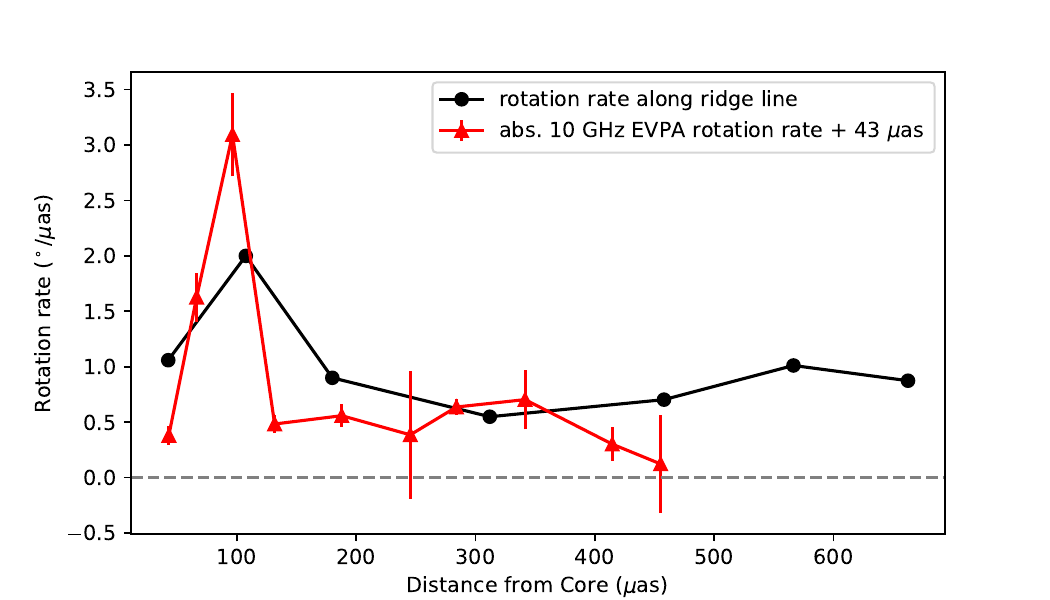}
\caption{Comparison of the rotation rate of the polarization angle rotation event at 10.45~GHz shown in \cite{2018A&A...619A..88M} and the rotation rate of the helical jet shape along the jet direction as seen in the 2016 RadioAstron observations of \object{OJ\,287}. In both cases we see a tighter winding of the jet helical, with it bending closer to the jet base.}
\label{fig:ridge_derivative}
\end{figure}

To quantitatively assess this relationship, we applied ridgeline analysis to our 2016 RadioAstron image. Assuming the ridgeline represents a projected helical trajectory, its resemblance with a sine wave of increasing period along the (mean) jet direction suggests that the step of the helical trajectory also increases gradually along that direction. We estimated the rotation rate of the helix along the (mean) jet direction by splitting the ridgeline in seven segments, each one corresponding to a quarter of the (increasing) period, which corresponds to a 90$^\circ$ change along the helix for each segment. The resultant helical trajectory rotation rate is shown as the black line in Fig.~\ref{fig:ridge_derivative}.

For comparison, the red line in Fig.~\ref{fig:ridge_derivative} represents the absolute 10\,GHz EVPA rotation rate reported in Fig.~4 of \cite{2018A&A...619A..88M}, converted from $^\circ$/day versus days to $^\circ$/$\mu$as versus $\mu$as. To change from temporal (days) to spacial ($\mu$as) scales, we assumed that the polarized component travels at a relativistic velocity and also moved from the observed to the emission reference frame using the mean Doppler factor and the redshift of the source as done also in Eq. 1 of \cite{2018A&A...619A..88M}. Finally, for the conversion of linear to angular distance we adopted 4.48 pc/mas, as expected at the redshift of OJ\,287.

The striking similarity between the two rotation rates in Fig.~\ref{fig:ridge_derivative} indicates that the polarized component responsible for EVPA rotation indeed traversed a more tightly wound jet bend initially, consistent with the increased ridgeline rotation rate near the jet base. Since the exact region of the jet responsible for the single-dish EVPA rotation might be different from the one depicted in the 2016 RadioAstron image, we arbitrarily shifted the EVPA rotation rate in Fig.~\ref{fig:ridge_derivative} by 43\,$\mu$as to align the two peaks and enable a direct comparison.

In summary, our high-resolution RadioAstron images from 2016 provide a cohesive picture that aligns with the high-cadence single-dish polarization monitoring data and the EVPA variability at the inner jet within the 43\,GHz core, as reported by \cite{2018A&A...619A..88M}.

\section{Summary and conclusions}
\label{sec:conclusion}

We have presented new, high-resolution RadioAstron observations of \object{OJ\,287} at 22\,GHz obtained on April 25, 2016. These observations, conducted at an angular resolution of $\sim47$\,$\mu$as, reveal complex jet dynamics on extreme spatial scales. Our main findings are as follows:

\begin{enumerate}
\item \textbf{Ribbon-like jet structure with multiple sharp bends.} 
Our space VLBI images reveal an exceptionally complex inner jet morphology in \object{OJ\,287} featuring three distinct bends within $\sim650$\,$\mu$as of the VLBI core; this is the first time such a morphology has been observed. This observation provides direct evidence that the full multi-scale reorientation of the jet occurs within the acceleration and collimation zone. The observed structure indicates a more complex jet evolution than previously assumed, possibly influenced by jet precession, magnetohydrodynamic instabilities, or interactions with the surrounding medium.

\item \textbf{Extreme brightness temperatures exceeding $10^{13}$\,K.}
Our model-fitting analysis identifies four key components: C1 (the VLBI core), C2 (further downstream), and B1 and B2 (located at bending points). The core exhibits an exceptionally high brightness temperature of $T_{\mathrm{b}} \sim 1.3 \times 10^{13}$\,K, exceeding the inverse Compton and equipartition limits. This implies strong Doppler boosting ($\delta \sim 10-30$), which is consistent with prior estimates of \object{OJ\,287}'s jet viewing angle ($\theta_\mathrm{j} \approx 3^\circ-8^\circ$). 

\item \textbf{Ordered magnetic field revealed by polarization structure.} 
Our polarimetric analysis shows EVPAs predominantly transverse to the jet axis in the core region, indicating a dominant poloidal magnetic field component. 

\item \textbf{Jet PA variations partially support precession models.}  
We tracked the evolution of the innermost jet PA using VLBA-BU-BLAZAR observations at 43\,GHz from 2014--2017. The jet shows a PA evolution of $\sim$30$^\circ$ over four years. These variations are partially consistent with precession models, particularly the disk precession model of \cite{Dey2018}, which aligns with some of our 43\,GHz data points. However, we note that not all epochs fit the predictions, and further observations would be necessary to draw firm conclusions. Therefore, while our data do not directly confirm the SMBHB scenario, they offer valuable structural and temporal constraints that can inform and refine such models.

\item \textbf{Connection between the jet's new components and high-energy flares.} 
Existing kinematic data at 43 and 86\,GHz combined with our 22\,GHz Gaussian model-fitting confirm the emergence of a new jet component (B15 or K) around March 2016, shortly after the December 2015 optical ``Centenary Flare.'' This component later interacted with a stationary recollimation shock (S1), and this interaction coincided with a TeV flare in early 2017. These findings further support a physical connection between disk impacts, jet activity, and high-energy emission in \object{OJ\,287}.

\item \textbf{Consistency between jet bending and EVPA rotation.}  
A comparison with the high-cadence single-dish polarization monitoring dataset presented in \cite{2018A&A...619A..88M} shows a remarkable agreement between the observed EVPA rotation and the jet ridgeline rotation rate. This suggests that the EVPA swings observed in 2016 trace an underlying helical bending of the jet.

\end{enumerate}

\begin{acknowledgements}

We thank Etienne Bonnassieux for his useful comments and fruitful discussions on the manuscript. 
Authors E. Traianou, and J. L. Gomez acknowledge financial support from the Severo Ochoa grant CEX2021-001131-S funded by MCIN/AEI/10.13039/501100011033. The work at the IAA-CSIC was supported in part by the Spanish Ministerio de Econom\'{i}a y Competitividad (grant number PID2022-140888NB-C21). 
I.C. is supported by the KASI-Yonsei Postdoctoral Fellowship program. This study makes use of VLBA data from the VLBA-BU Blazar Monitoring Program (BEAM-ME and VLBA-BU-BLAZAR;
\url{http://www.bu.edu/blazars/BEAM-ME.html}), funded by NASA through the Fermi Guest Investigator grants, the latest is 80NSSC23K1508. The RadioAstron project is led by the Astro Space Center of the Lebedev Physical Institute of the Russian Academy of Sciences and the Lavochkin Scientific and Production Association under a contract with the State Space Corporation ROSCOSMOS, in collaboration with partner organizations in Russia and other countries. The VLBA is an instrument of the National Radio Astronomy Observatory. The National Radio Astronomy Observatory is a facility of the National Science Foundation operated by Associated Universities, Inc. This research has made also use of data obtained with the Global Millimeter VLBI Array (GMVA), which consists of telescopes operated by the MPIfR, IRAM, Onsala, Mets\"ahovi, Yebes, the Korean VLBI Network, the Green Bank Observatory and the Very Long Baseline Array (VLBA). The VLBA is a facility of the National Science Foundation operated under cooperative agreement by Associated Universities, Inc. The data were correlated at the special version \citep{2016Galax...4...55B} of the RadioAstron \texttt{DiFX} correlator \citep{2011PASP..123..275D} of the MPIfR in Bonn, Germany. 

This research has made use of NASA's Astrophysics Data System.

MW is supported by a~Ram\'{o}n y Cajal grant RYC2023-042988-I from the Spanish Ministry of Science and Innovation.

LIG gratefully acknowledges support by the Chinese Academy of Sciences PIFI program, grant No.~2024PVA0008. 

ABP is supported in the framework of the State project ``Science'' by the Ministry of Science and Higher Education of the Russian Federation under the contract 075-15-2024-541.

MML, YYK, and APL were supported by the M2FINDERS project that had received funding from the European Research Council (ERC) under the European Union's Horizon 2020 research and innovation programme (grant agreement No~101018682).

This work is partly based on observations carried out with the IRAM 30m telescope. IRAM is supported by INSU/CNRS (France), MPG (Germany) and IGN (Spain).

The research at Boston University was supported in part by NASA Fermi GI grant 80NSSC20K1567.

\end{acknowledgements}

\bibliographystyle{aa} 
\bibliography{aanda}
\clearpage
\begin{appendix}

\section{Image fidelity and data issues}
\label{appA}

During the data processing, we flagged multiple stations and IFs due to poor data quality. Specifically, stations KT, KU, BD, and RO were excluded because of the absence of fringes, while the first and fourth IFs of station ON were flagged for corrupted signals. We used GB as the reference antenna throughout the analysis. Station LA displayed erratic amplitude fluctuations despite having stable phases. Therefore, we adopted a special strategy: LA was included in closure-phase imaging but excluded from amplitude-based imaging, and it was used only during self-calibration steps, ensuring that LA amplitudes followed the more robustly constrained model derived from the other antennas.

\begin{figure}
\centering
\includegraphics[width=0.455\textwidth]{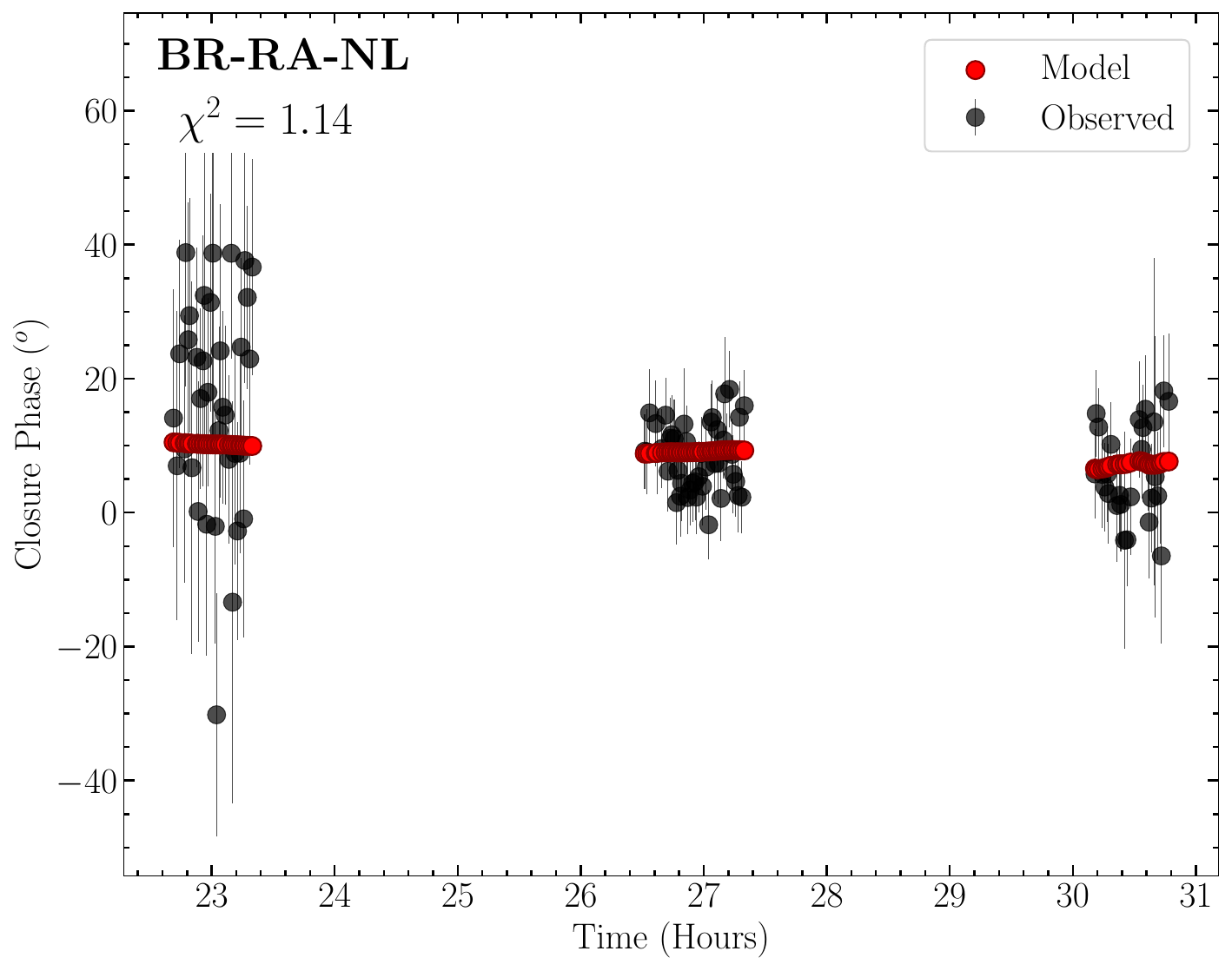}
\includegraphics[width=0.455\textwidth]{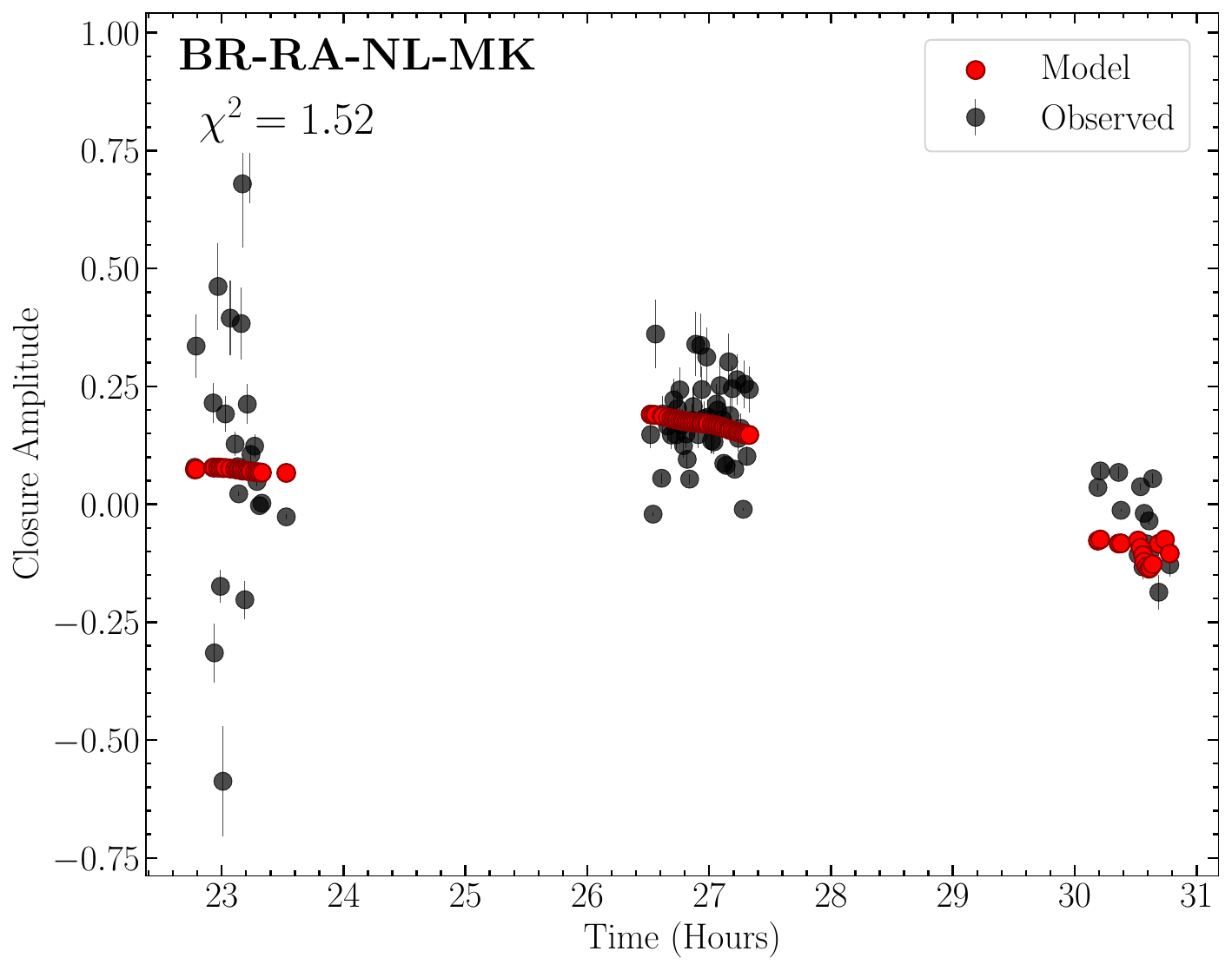}
\caption{Selected closure amplitudes and phases from coherently averaged visibilities on triangles as a function of time.}
\label{fig:closures}
\end{figure}

\begin{figure}
\centering
\includegraphics[width=0.455\textwidth]{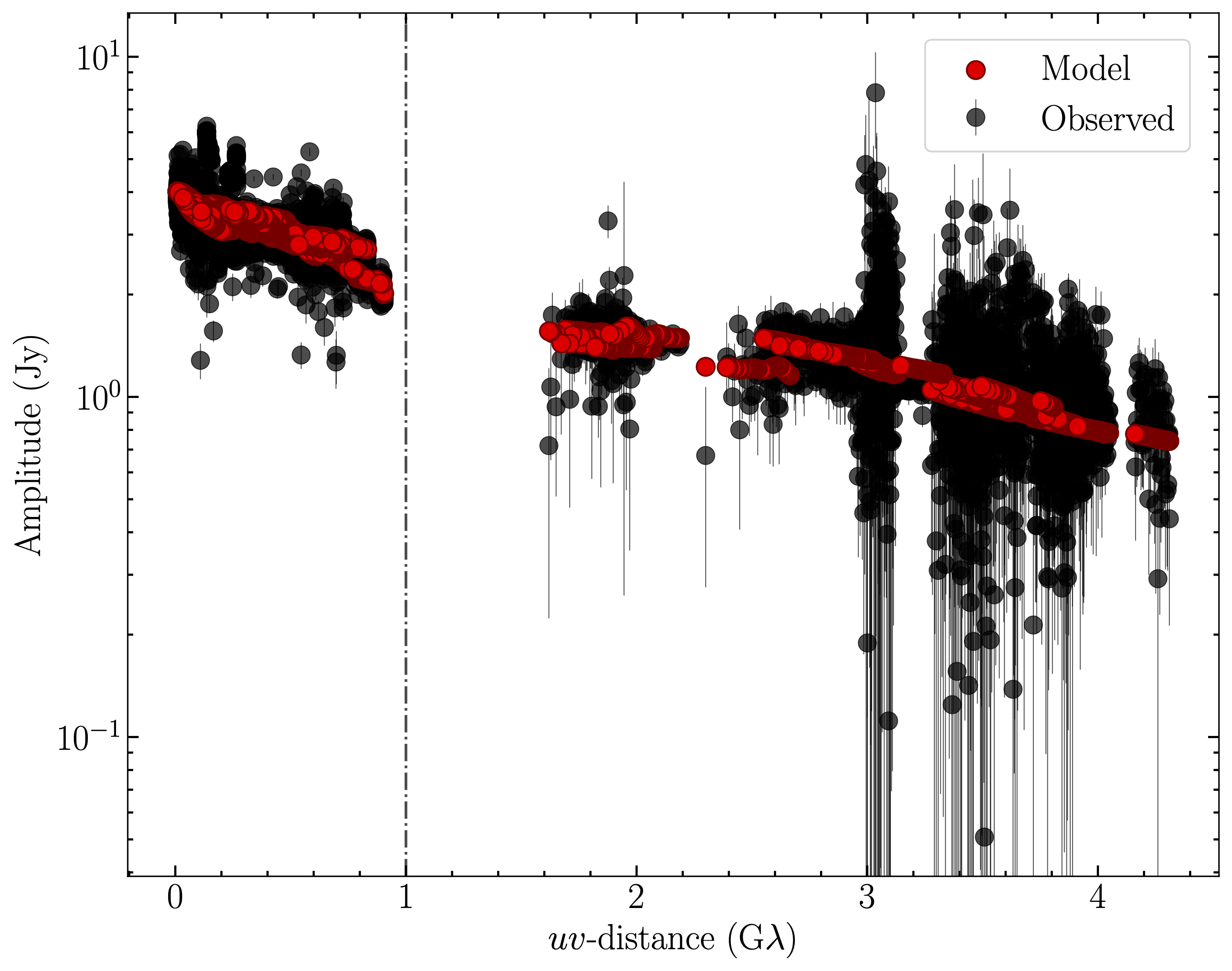}
\includegraphics[width=0.455\textwidth]{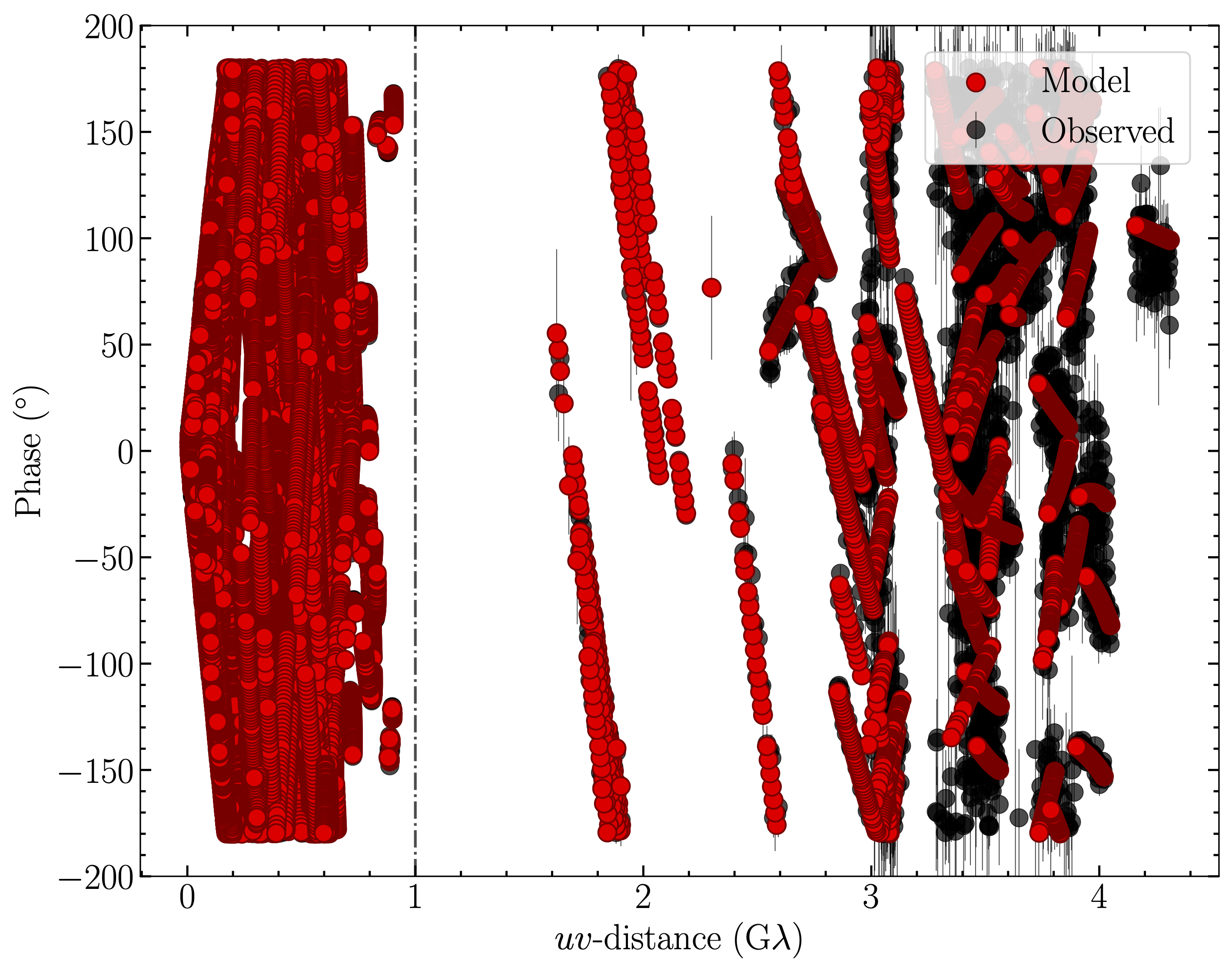}
\caption{Self-calibrated visibility amplitudes and phases as a function of uv distance from the RadioAstron observations of \object{OJ\,287 from} April 24--25, 2016, at 22\,GHz. The magenta points represent the fit to the data using the \texttt{ehtim} model derived from the imaging process. Reliable space-ground fringe detections were achieved up to a projected baseline length of 4.6 Earth diameters. The dot-dashed  black line separates the ground-only baselines from those involving the SRT.}
\label{fig:radplot}
\end{figure}

\begin{figure}
\centering
\includegraphics[width=0.5\textwidth]{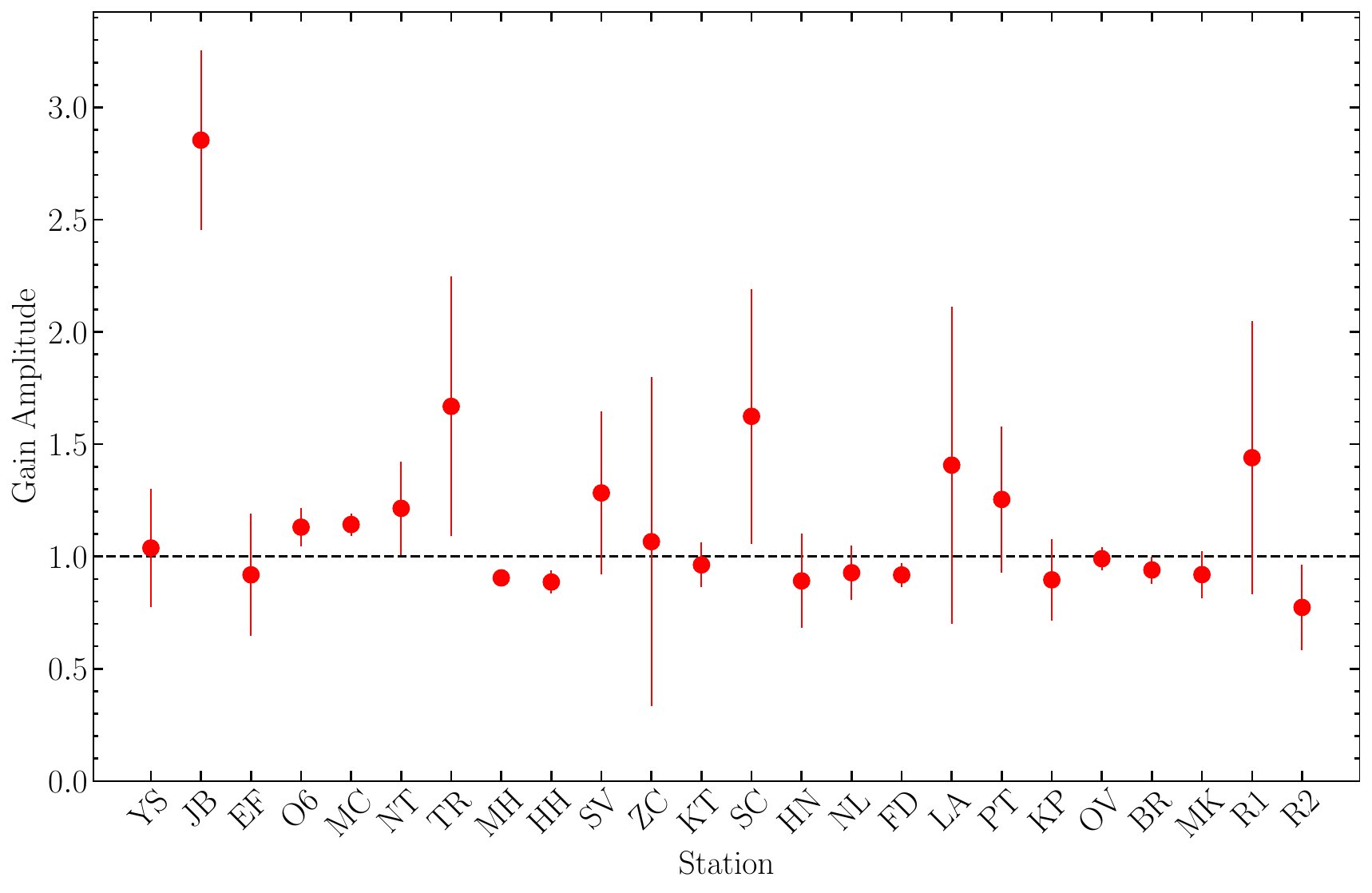}
\caption{Multiplicative gain correction factors at each station.}
\label{fig:gains}
\end{figure}

\FloatBarrier

\section{Instrumental polarization}
\label{appB}

We note that station JB was flagged due to an excessively large gain value of \(2.85 \pm 0.4\). For polarization imaging, we flagged additional stations HH, KT, MK, NT, PT, EF, and SR because of insufficient parallactic angle coverage or unstable signal fluctuations (particularly noted at PT). The D-term calibration for RadioAstron yielded \((-0.0174 + 0.0111i)\) for RCP and \((0.0198 - 0.0001i)\) for LCP.

\begin{figure}
\centering
\includegraphics[width=0.455\textwidth]{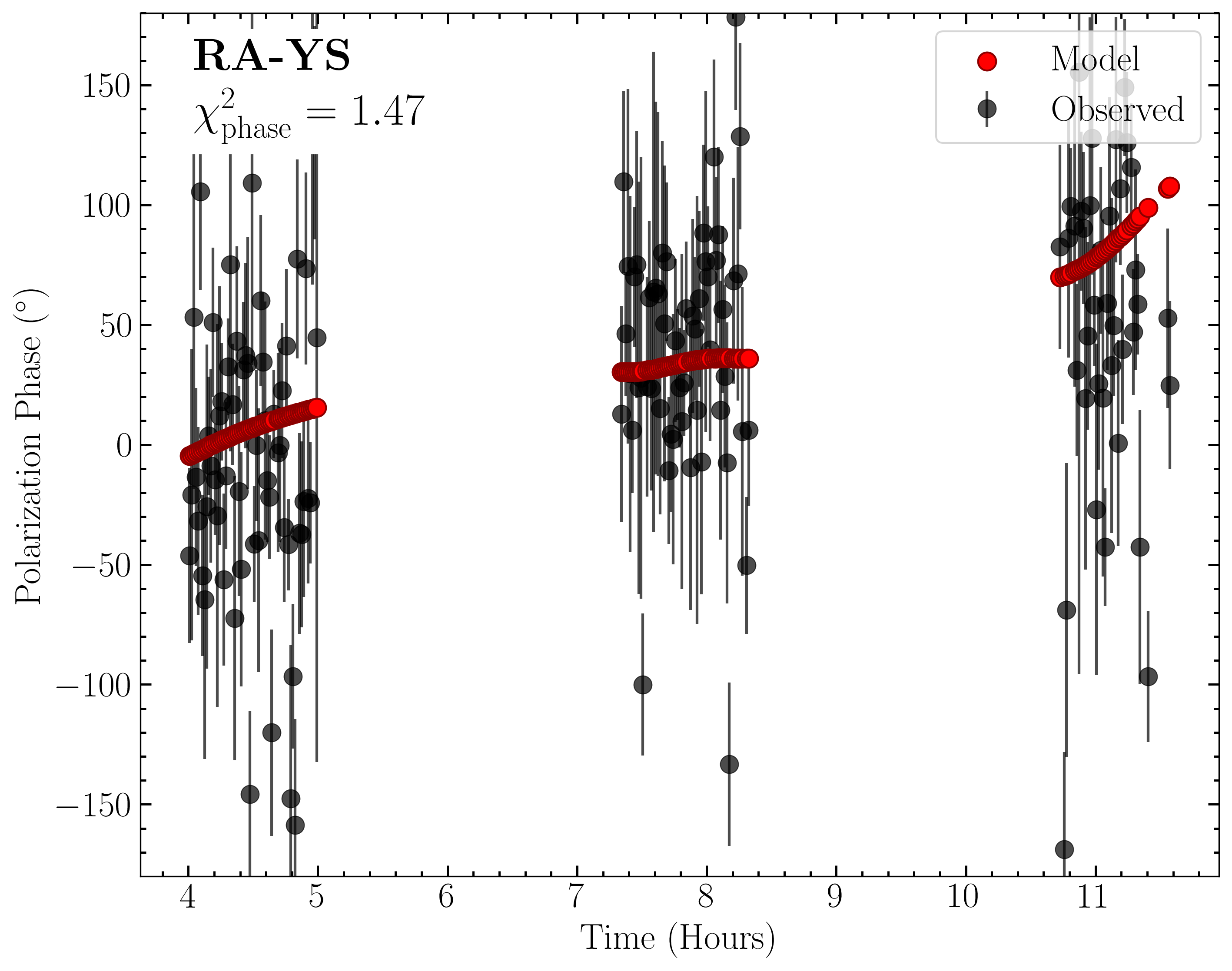}
\includegraphics[width=0.455\textwidth]{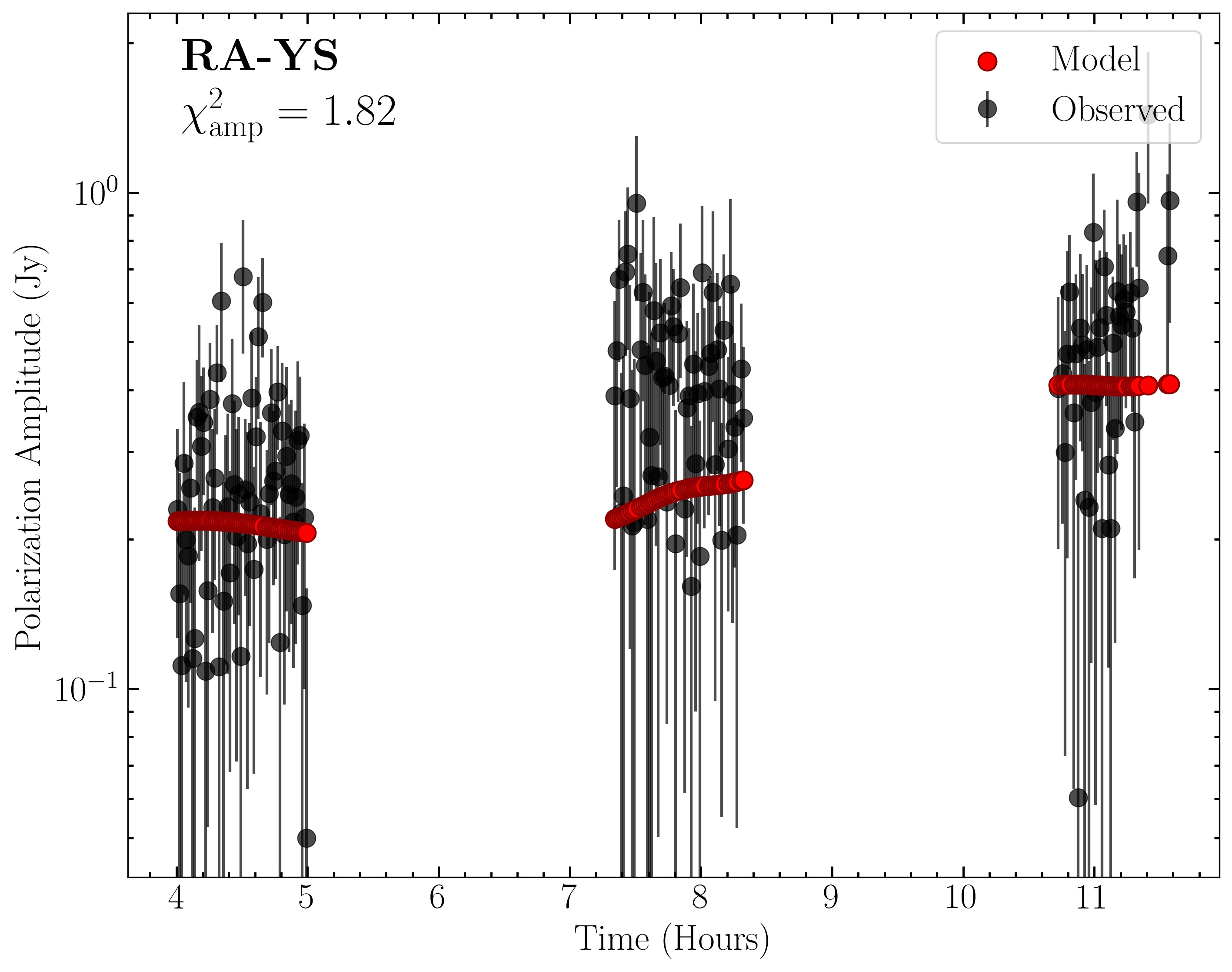}
\caption{Time evolution of selected baseline polarization amplitudes and phases derived from coherently averaged visibility measurements.}
\label{fig:polbaselines}
\end{figure}

\begin{figure}
\centering
\includegraphics[width=0.5\textwidth]{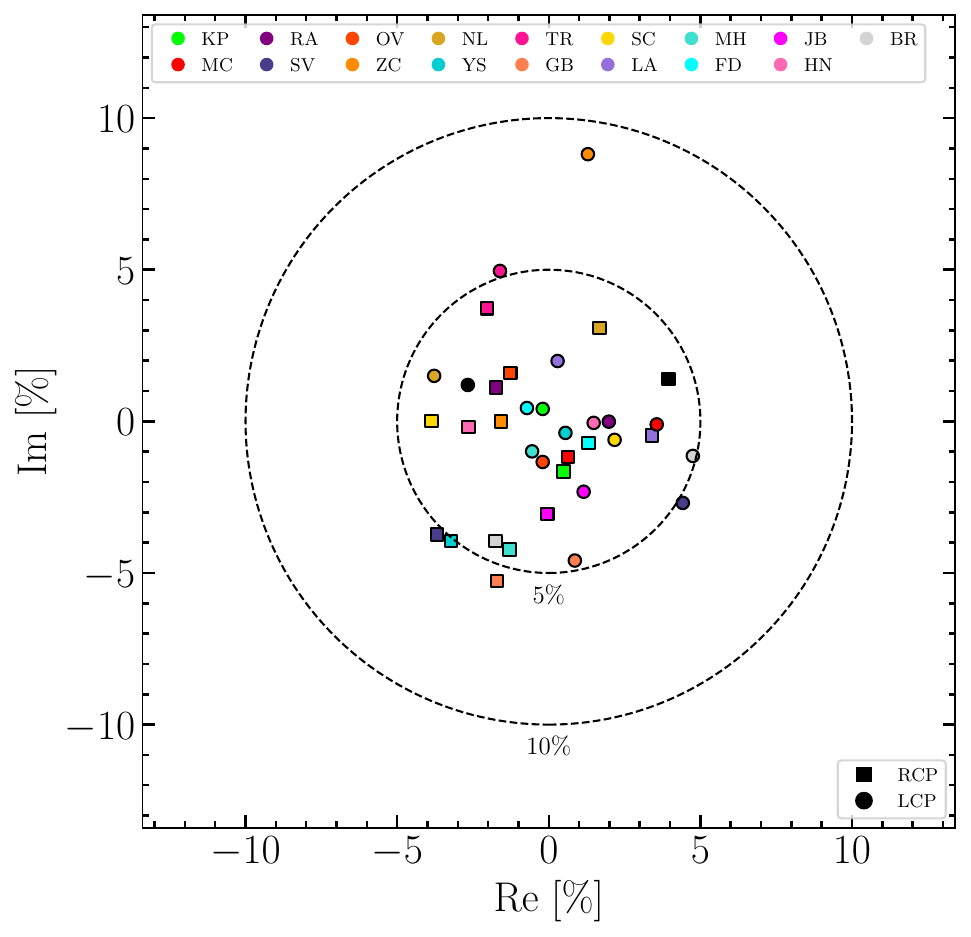}
\caption{D-term solutions for each station and each polarization. Labels mark the different radio telescopes.}
\label{fig:dterms}
\end{figure}

\end{appendix}


\end{document}